\documentclass{elsart}
\newcommand{\fref}[1]{\mbox{Fig.$\!$~\ref{#1}}}
\input epsf
\begin{document}
\begin{frontmatter}
\title{Calculations of fission rates  for r-process
nucleosynthesis.
\thanksref{Someone}
 }
\author[basel,moscow]{I.V.Panov\thanksref{X}}
\author[basel]{E.Kolbe}
\author[mainz]{B.Pfeiffer}
\author[basel]{T.Rauscher}
\author[mainz]{K.-L.Kratz}
\author[basel]{F.-K. Thielemann }
\address[basel]{ University of Basel, Klingelbergstr. 82,
  CH-4056 Basel,     Switzerland}
\address[moscow]{Institute for Theoretical and Experimental
 Physics,  B.Cheremushkinskaya 25, Moscow, 117259, Russia.}
\address[mainz]{Institute fur Kernchemie,
Fritz-Strassman-Weg 2, D-55128 Mainz,    Germany}
\thanks[Someone]{Partially supported by Swiss National Fonds}
\thanks[X]{corresponding author: Igor.Panov@itep.ru }

\begin{abstract} Fission plays an important role in the r-process which is
responsible not only for the  yields  of transuranium isotopes,
but may have a strong influence on the  formation  of the majority
of heavy nuclei due to fission recycling. We present calculations
of beta-delayed and neutron-induced fission rates, taking into
account different fission barriers predictions and mass formulae.
  It is shown that an increase  of fission barriers results
  naturally in a reduction of  fission rates,
  but that nevertheless fission leads to the termination of the r-process.
  Furthermore, it is discussed that the probability of triple fission could be
  high for $A>260$ and have an effect on the
  formation of the abundances of heavy nuclei.  Fission
  after beta-delayed neutron emission is discussed as well as
  different aspects of the influence of fission upon r-process
  calculations.

\end{abstract}
\begin{keyword}
fission; fission rates;  r-process;

\PACS 21.10.Dr; 24.60.Dr; 24.75.+i; 25.85.Ec; 25.85.Ca; 26.30.+k

\end{keyword}
\end{frontmatter}

\section{Introduction}

 Fission is an important process which is responsible not only for
the yields of the transuranium isotopes in the r-process
calculation and for termination of the r-process, but also for the
formation of the majority of heavy nuclei  due to cycling of the
r-process from transuranium region into the region of
A$\approx$130 (where fission products are involved again in the
r-process nucleosynthesis). For a consistent treatment of the
r-process, we have to consider fission barriers based on the same
mass formulae that are used for the calculation of other reaction
rates and decay properties. We have to evaluate all fission
branches: neutron-induced fission, beta-delayed fission and
spontaneous fission. In addition, one should analyze the
importance of every separate fission mode for the termination of
the r-process, as well as nuclear abundance yields and the related
age determination from the chronometer nuclei abundances.

The fission barriers, used in the past for fission rate
calculations \cite{homo80} were probably underestimated. Recent
calculations \cite{mysw99,mamdo98} predict higher fission
barriers, sometimes significantly. Therefore, a recalculation of
fission rates is important as well as an evaluation of their
sensitivity on nuclear data.
      We consider all fission channels, but first discuss
beta-delayed and neutron-induced fission.

While the study of fission in the r-process has a long history
\cite{SFC65},
its main focus was related to cosmochronometers
\cite{{ctt87},{cowan99},{Chech88}}.
%
Only chronometer abundances are affected if the neutron freeze-out
occurs when the r-process  (in the envelope of a supernova)
approaches the transuranium region and yields of fission products
play small role for the abundances of all light r-process nuclei.
For high neutron densities existing over long periods, fission
cycling to lighter nuclei becomes important. For the latter case,
considering only instantaneous fission is a first step towards a
consistent treatment as performed in a number of recent studies:
   (a) $ P_{fission} \equiv P_{sf}=1$ for all A $>$ 240
\cite{frei99}; \hspace{5mm}
   (b) $ P_{fission} \equiv P_{sf}=1$ for all A $>$ 256
\cite{cowan99};
   (c)    for A $=$ 260 \cite{rausch94}. In all these cases only
a symmetric mass distribution was considered, while other
calculations already included asymmetric mass distributions
\cite{{SFC65},{Chech88}}.

 In the first extended calculation of beta-delayed fission for
 the majority of r-process nuclei \cite{TMK83}, it
turned out that the fission probability could reach 100\% and
beta-delayed fission affects strongly the transuranium yields,
especially the yields of cosmochronometer nuclei. Their yields are
used to determine the stellar age or the age of our galaxy as the
lower limit on the age of the universe.
     For r-process conditions with long durations of high neutron
densities, neutron-induced fission could also play an important
role. That can be seen from the comparison of neutron-induced
fission rates for uranium isotopes \cite{tcc89_50} employing the
old barrier estimates \cite{homo80} and extended calculations for
a variety of fission barrier models \cite{Paz03}.
     The availability of new extended fission barrier calculations
\cite{mysw99,mamdo01}, requires that all the fission rates need to
be reevaluated for the inclusion in r-process calculations.

 The latest observations of metal-poor stars emphasize also the
importance of the mass region $110<A<130$ for the understanding of
the role of fission in the r-process. One notices that these
nuclei are underabundant and the solar system r-process must have
at least two components, the one dominating for A$<$130 and the
other one dominating for A$>$130 \cite{sneden00}. The observation
of the second component in low metallicity stars and the observed
abundances for A$<$130 might  give strong clues to fission
properties, if the explanation of the second component is related
to an r-process site with a high neutron supply which leads to
fission cycling. This should be investigated as the site of the
r-process is still uncertain.

 Here we want to emphasize that, besides
neutron-induced fission and beta-delayed fission rates, the
mass-distribution  of fission yields is important as well as it
was discussed earlier \cite{pft00}. Although the mass distribution
of fission products is known rather well for long-lived
transactinide nuclei, the mass distribution after fission of
neutron-rich heavy elements is an open question. The r-process
yields in the mass region of $80<A<120$ can be affected by the
mass distribution of fissioning nuclei, and triple or ternary
\cite{ternmas} fission can play an important role in some cases.

Beta-delayed fission rates differ very strongly due to
difficulties to predict reliable beta-strength functions for very
neutron-rich and especially for deformed nuclei as well as
reliable fission barriers for such nuclei.

\begin{figure*}
\centerline{ \epsfxsize=0.65\textwidth\epsffile{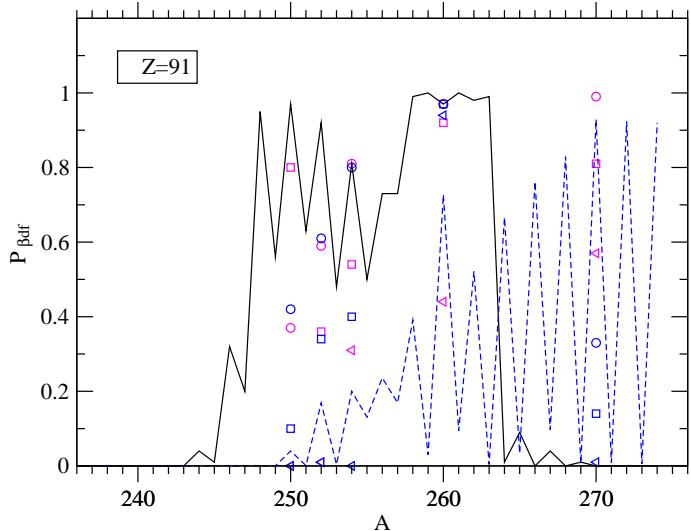}}
  \caption{The calculated probabilities for beta-delayed fission $P_{bdf}$
for Pa- isotopes from different calculations using approach of
Tamm-Dancoff \cite{TMK83} (line), QRPA \cite{staud92} (dashed
line) and RPA approximation with some artificial parametrization
\cite{meyer89} (circles, squares and triangles).
 }
  \label{pbdfold}
\end{figure*}

\begin{figure*}
\centerline{
\hspace{-2mm}\epsfxsize=0.52\textwidth\epsffile{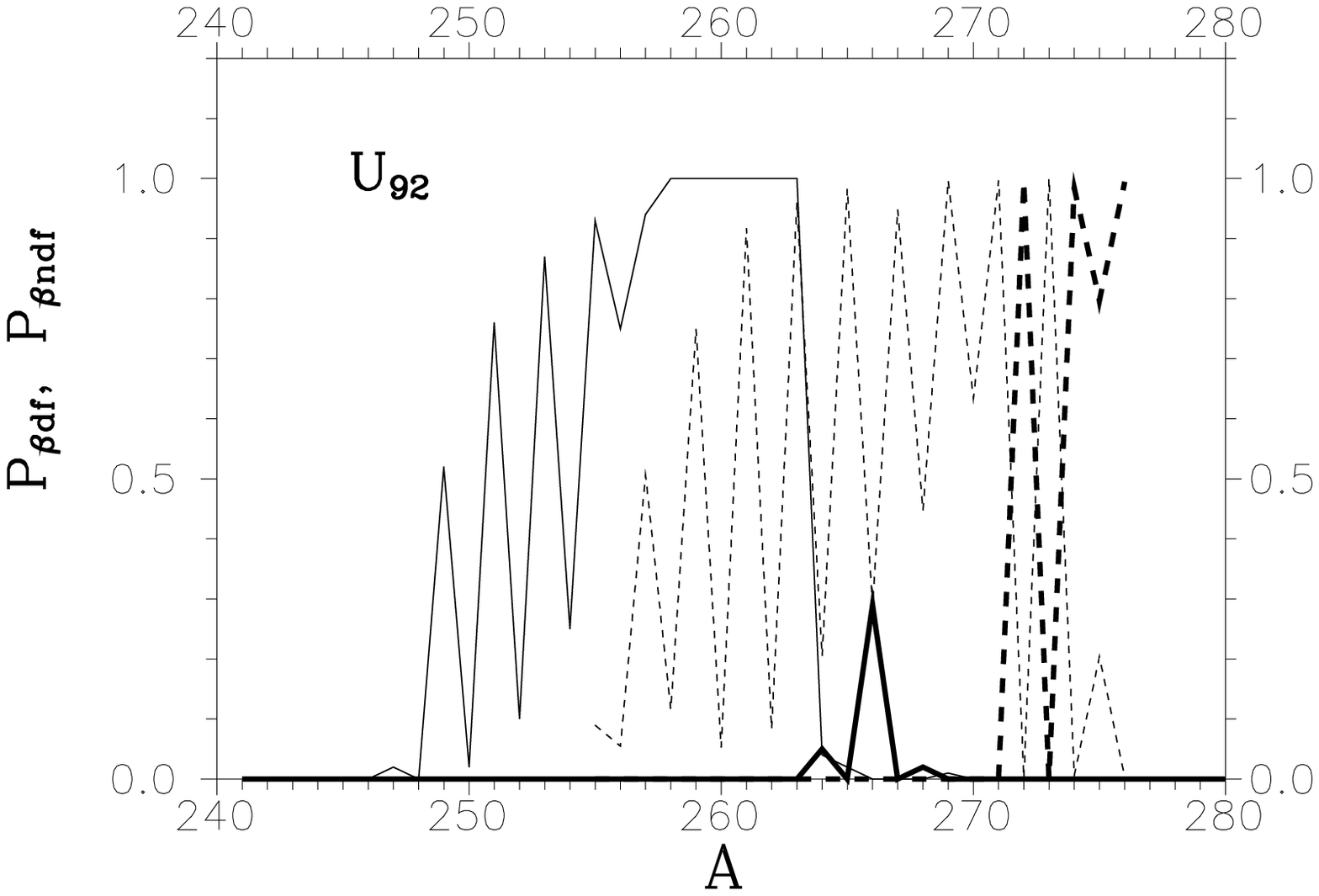}
\hspace{2mm}\epsfxsize=0.52\textwidth\epsffile{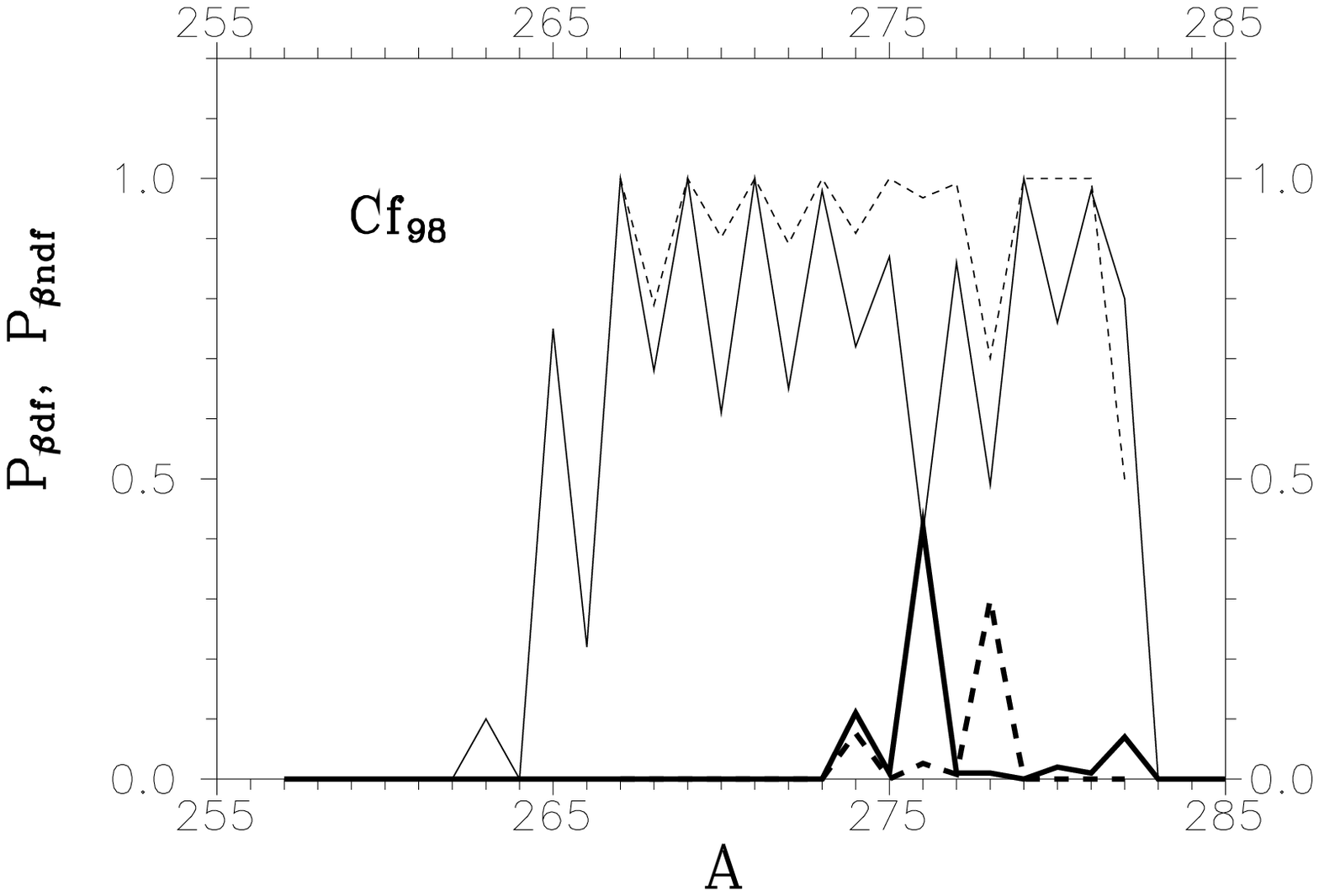}}
\vspace{2mm}
  \caption{Fission probabilities $P_{\beta df}$ (thin lines) and $P_{\beta
  ndf}$ (bold lines)
 for  U  and Cf  isotopes for different
strength-functions: Tamm-Dankoff \cite{TMK83} (lines) and pn-QRPA
\cite{staud92} (dashed lines). Mass predictions \cite{hilf76} and
fission barriers of ref. \cite{homo80} were used. }
  \label{tmk83st92}
\end{figure*}

 In a number of calculations \cite{{TMK83},{staud92}},
 (contrary to \cite{PLL90}) the values of beta-delayed fission
  probabilities for a large number of   short-lived neutron-rich
  heavy nuclei can attain 100\%, when
applying the fission barriers of  \cite{homo80}, (especially for
transuranium nuclei) independent of beta-strength functions were
used (see \fref{pbdfold}, \fref{tmk83st92}).
    The beta-strength function $S_{\beta}$ used
in \cite{TMK83} was based on Tamm-Dancoff approximation for the
Gamow-Teller residual interaction.
    Another calculation  \cite{staud92} used the
proton-neutron Quasiparticle Random Phase Approximation (pn-QRPA)
\cite{pnqrpa} with a shematic Gamow-Teller residual interaction.
Three sets of calculations with 3 different mass formulae
 \cite{homo80,groot76,hilf76} were considered taking into
 account an implicit treatment of nuclear deformation.

Beta-delayed fission calculations for a number of selected nuclei
 were also performed   \cite{meyer89} with a
schematic consideration of deformations as in \cite{staud92}, to
mix a basis of deformed nuclear states via the GT residual
interaction.
           \fref{pbdfold} shows some values of
$P_{\beta df}$   with different parameterization of the
$\beta$-strength-function (squares) and artificial overestimation
(triangles) and underestimation (circles) of fission barriers and
values from existing data sets \cite{{TMK83,staud92}}  for the
isotopes of Pa (Z=91).
         Both of the existing data sets
\cite{{TMK83,staud92}} strongly depend on the $\beta$-strength
function, mass models and fission barrier heights, and their
application to r-process models gave quite different results
\cite{pft00}.

      The degree  of difference depends on the r-process
model used. In the case of the classical model of the r-process,
the path of the r-process  lies along nuclei with neutron
separation energy $S_n \approx$ 2MeV, and the r-process stops at
mass numbers A$\approx$260. Both of the data sets
\cite{TMK83,staud92} yield the maximum values of $P_{\beta f}$ for
these nuclei. That is why the previously used approximation of
100\% instantaneous fission in the vicinity of A$\sim$260
\cite{cowan99}, \cite{rausch94}, \cite{tcc89_50} was a fairly good
approximation.
    But when considering extreme r-process paths in the framework of
the classical model ($S_n\sim$ 3 MeV or $S_n\sim$ 1 MeV) or with a
calculation of the heavy element abundances in a very high neutron
density environment, the difference in the results with
utilization of different beta-delayed fission rates can be large.

It has to be noted that for very neutron rich nuclei the
possibility of fission after delayed neutron emission could be
rather high and, that in principle, such a fission mode also
should be calculated.
        In  \fref{tmk83st92}, different calculations of the beta-delayed
fission probabilities \cite{TMK83,staud92}  with the same fission
barriers \cite{homo80} are compared, including the rates of
fission after beta-delayed neutron emission. The strength-function
dependence and the high values of $P_{n \beta df}$ are clearly
seen, especially for heavy Cf-isotopes.

  The details of the statistical approach will be introduced in
Section 2, including a discussion  of the beta-strength function
and probability of fission after beta-delayed neutron emission. A
comparison of different fission barriers and mass predictions used
in present calculations is shown in Section 3.
      In Section 4 we present the results. The impact analysis and
a comparison of the different decay modes in fission is discussed
in subsection 4.1 with evaluations based on older mass predictions
\cite{hilf76} and fission barriers \cite{homo80}, often used in
astrophysical applications.  In subsection 4.2, we discuss  the
fission rates for more recent nuclear physics input from the
Thomas-Fermi \cite{mysw99} and ETFSI \cite{mamdo01} models and
discuss the changes. In Sections 5 and 6, the possible
consequences of fission rates are considered: cycling of the
r-process due to fission into the region of A$\approx$130, and the
influence of triple fission on the yields of medium mass nuclei.

\section{The statistical approach to fission rates}

As in previous approaches \cite{TMK83,staud92}, we have applied
the statistical Hauser-Feshbach formalism for the calculation of
fission rates. It has been shown  \cite{RTK97} that the
statistical model is very well applicable for the astrophysical
neutron-induced rate calculations, as long as there is a high
density of excited states, which is the case for heavy nuclei. Of
course, just near the neutron drip-line the  systematic errors of
the approach in neutron-induced fission rate calculations can
rise, underlining that reliable mass predictions are absolutely
necessary for r-process applications far from stability. Early
r-process calculations made use of the mass predictions by Hilf et
al. \cite{hilf76}  and the fission barriers of ref. \cite{homo80}
for fission rate evaluations. For a consistent treatment of
fission rates, however, the neutron separation energies, reaction
Q-values and fission barrier heights should be used from the same
mass model.

\subsection{Neutron-Induced fission and beta-delayed fission}

The cross-section for a neutron-induced reaction $i^0(n,f)$ from
the target ground state $i^0$ with center of mass energy $E_{in}$
and reduced mass $\mu_{in}$ is given by

\begin{equation}  \hspace{-10mm}
\sigma_{n f}(E_{in}) =
   \frac{\pi \hbar^2/(2\mu_{in}E_{in})}{  (2J^{0}_i+1)\cdot(2J_n+1)}
   \sum_{J,\pi} (2J+1) \frac{T^{0}_n (E,J^{\pi}, E^0_i,J^0_i,{\pi}^0_i)
   T_{f} (E,J^{\pi}) }
   { T_{\mathrm{tot}}(E,J^{\pi})}.
  \label{signhaus}
\end{equation}

 The transition coefficient $T_f(E,J^{\pi})$ includes the sum over all
possible final states. Since the work of Strutinski \cite{strut}
fission has been generally described within the framework of
double-humped fission barriers. Similar to previous work
\cite{TMK83} we followed this approximation of a two-hump barrier.
The calculation of the fission probabilities was performed in the
complete damping approximation which averages over transmission
resonances, assuming that levels in the second minimum are equally
spaced.

Then beta-delayed fission rate can be expressed by
\begin{eqnarray} \lambda_{f}=\int^{Q_{\beta}}_0 \sum_i
\beta_i(E,J_i,\pi_i)\cdot
\frac{T_f(E,J_i,\pi_i)}{T_{\mathrm{tot}}(E,J_i,\pi_i)} dE   
\end{eqnarray}

where  the beta-feeds $\beta_i(E,J_i,\pi_i)$ are expressed via
reduced transition probabilities $B(E,J_i,\pi_i)$, taking into
account a Gauss spreading of states:
\begin{eqnarray}
\beta_i(E,J_i,\pi_i)=\frac{1}{\sigma_i(2\pi)^{1/2}}
exp[-(E-E_i)^2/2\sigma^2_i] B(E,J_i,\pi_i)f(Q_{\beta}-E)
\end{eqnarray}

with integrated Fermi function
\begin{eqnarray}f(Q_{\beta}-E)=\int_1^{E_0}F(Z,\epsilon)\epsilon
\sqrt{(\epsilon^2-1)} (E_0-\epsilon)^2d\epsilon .  \nonumber
\end{eqnarray}
$T_{f}(E,J_i,\pi_i)/T_{\mathrm{tot}}$ denotes the fission
probability of the compound nucleus:

\begin{eqnarray}
\frac{T_f(E,J_i,\pi_i)}{T_{\mathrm{tot}}(E,J_i,\pi_i)}=
  \{ 1+ \left(\frac{T_n(E,J_i,\pi_i)+T_{\gamma}(E,J_i,\pi_i)}{
T_{\mathrm{eff}}(E,J_i,\pi_i)} \right)^2+
\end{eqnarray}
\begin{eqnarray} \hspace{-10mm}
 2\cdot
\frac{T_n(E,J_i,\pi_i)+T_{\gamma}(E,J_i,\pi_i)}{T_{eff}(E,J_i,\pi_i)}
 \cdot \coth[\frac{T_A(E,J_i,\pi_i)+T_B(E,J_i,\pi_i)}{2}] \}
^{-\frac{1}{2}} \; \nonumber
\end{eqnarray}

where
\begin{eqnarray}
T_{\mathrm{eff}}=\frac{T_A(E,J_i,\pi_i) \cdot
T_B(E,J_i,\pi_i)}{T_A(E,J_i,\pi_i)+T_B(E,J_i,\pi_i)}   \; , \
\nonumber
\end{eqnarray}

with \begin{eqnarray}
 \hspace{-1cm}T_{A,B}(E,J,\pi)=\int_0^E \rho_{A,B} (\epsilon, J, \pi)
T_{HW} ( E-E_{A,B} - \epsilon -h^2l(l+1)/2\theta, h\omega_{A,B} )d
\epsilon \; . \; \; \;  \;
\end{eqnarray}

$T_{HW}$ denotes the Hill-Wheeler \cite{HillW} transmission
coefficient through a parabolic barrier
\begin{eqnarray}
T_{HW}(E, h\omega) = \frac{1}{1+ exp(-2 \pi E/h\omega)} \; \; \; ,
\end{eqnarray}

and the available energy is reduced by the rotational energy with
the moment of inertia $\theta$ deduced from the irrotational flow
model \cite{BorM}. The fission barrier heights $E_{A,B}$ will be
discussed in Section 3. $\rho_{A,B}$ define the level densities
above the first and second saddle points, which show an
enhancement over the level densities at ground state deformation,
due to increased deformation and coupling to low-lying rotational
excitations. This enhancement is larger for the first axially
asymmetric barrier than for the second mass asymmetric barrier. As
in previous calculations, we used constant enhancement factors 4
or 2, respectively, over the level density at ground state
deformation
 \cite{{IAEA_smir},{bjlynn80}}.

The back-shifted Fermi-gas description of the level density was
improved by introducing an energy dependent level density
parameter {\it a} as described in ref.\cite{RTK97}. Tests show a
very good agreement with shell model calculations \cite{Dean} and
justify the application at and above the neutron separation
energy.

The amount of experimentally known fission barriers used in the
present calculations  was increased recently by approximately a
factor of 2
 following the IAEA-report \cite{IAEA_smir} and recent compilation in \cite{mamdo01}.
%

The relative probability of beta-delayed fission over pure
beta-decay is    $P_{\beta df} =\lambda_{f}/ \lambda $
 where $\lambda$ denotes the $\beta$-decay rate, obtained from
Eq. (5) without the term $T_f/T_{\mathrm{tot}}$.

\subsection{Beta-strength function }

The beta-strength functions, defined as

$$ S_{\beta}(E) =
\sum\limits_i \beta_i(E,J_i,\pi) $$

for our calculations were derived on the basis of the QRPA
calculations  of reduced transition probabilities $B(E,J_i,\pi_i)$
\cite{MoPfK02} (see formula (3)).

In our  calculations of beta-delayed fission probabilities we used
the new approach for the beta-strength function \cite{MoPfK02},
based on the approach \cite{MoRand90}, considered allowed
transitions and  deformed single-particle model as in
\cite{krumo84} and takes into account the residual interaction.
The model \cite{MoPfK02} combines calculations within the
quasi-particle RPA for the Gamow-Teller part with an empirical
spreading of quasi-particle strength and the gross theory for the
first-forbidden part of $\beta$-decay.


In total fission barriers, penetrabilities/widths of all competing
branches of the compound nucleus after beta-decay, and the  beta
strength functions are required to beta-delayed fission.

 $P_{\beta df}=\lambda_f/\lambda $
 using  Eq. (2) can be written  as

\begin{eqnarray}
%
   P_{\beta df}(Z,A)  =
\frac
  { \int\limits_0^{Qb}  S_{\beta}(E) \frac{T_{f}}{T_{\mathrm{tot}}}dE}
        { \int\limits_0^{Q_b}  S_{\beta}(E) dE}
\end{eqnarray}

\begin{figure*}
 \centerline{
\hspace{-7mm}\epsfxsize=0.85\textwidth\epsffile{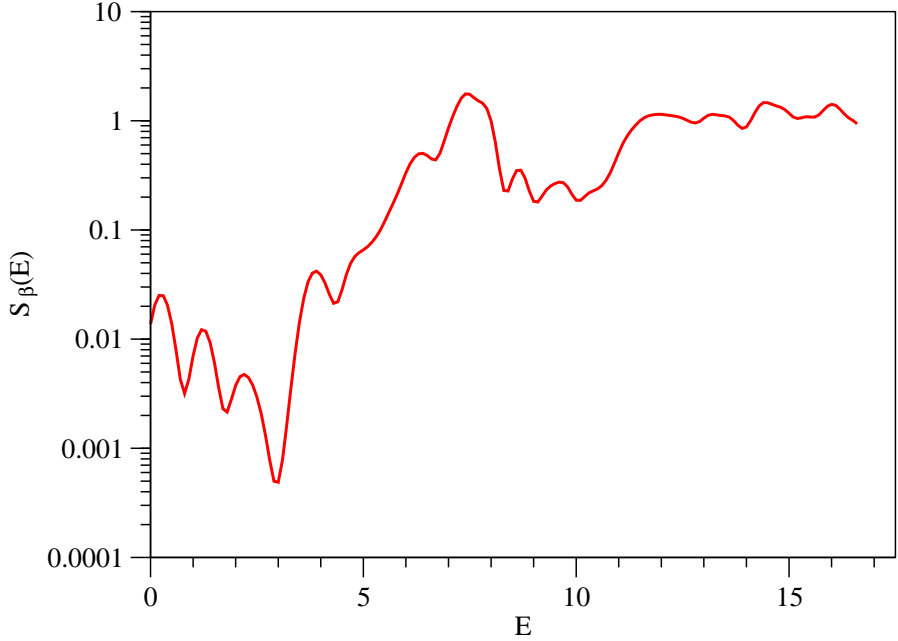} }
  \caption{Theoretical   $S_{\beta}(E)$ for $^{300}U$
}
  \label{sbet}
\end{figure*}


 To calculate $\beta$-decay Q-values and neutron
separation energies the experimental ground-state masses were used
where available, otherwise calculated masses (ETFSI
\cite{PearG96}, Thomas-Fermi \cite{mysw99} or  Howard and M\"oller
\cite{homo80}). To reduce the  uncertainties in the beta-strength
function calculation due to uncertainties in individual level
energies,   a spreading of individual resonances were introduced
above E=2 MeV.
       Specifically, each
"spike" in the beta-strength function above 2 MeV was transformed
into a Gaussian with width $\approx$ 0.5. The calculations were
based on  ground state deformations which affect the energy levels
and wave-functions that are obtained in the single-particle model.
The ground-state deformations were calculated in the FRDM mass
model \cite{mol92}. In some cases the first-forbidden strength
calculated in the gross theory \cite{kota75}  was included.

       The example of beta-strength function for the decay of $^{300}U$ is shown in (\fref{sbet}).

\subsection{Cascade Model for secondary fission}

All heavy nuclei near the r-process path are very neutron rich.
Therefore the neutron separation energies for the daughter nuclei,
produced in beta-decay,  are very low. As a result, the
probability of beta-delayed neutron emission can be high,
frequently higher than probability of beta-delayed fission,
(depending on fission barrier value). Therefore the excited states
of a daughter nucleus produced by the beta-decay of neutron rich
transuranium nuclei, mainly decay by neutron emission or fission.
If neutron emission is the dominant decay mode it can lead to
excited states that lie above the respective fission barrier and
fission could occur in the second decay-step. This probability of
secondary fission - $P_{\beta dnf}$, i.e., fission of the compound
nucleus that is produced by neutron emission of the excited state
of the daughter nucleus, will contribute to the total fission
probability $P^{tot}_{\beta df}$ and might even enhance it
significantly:

\begin{eqnarray}
%
  P^{\mathrm{tot}}_{\beta df}(Z,A)=P_{\beta df}(Z,A)  + P_{\beta dnf}(Z,A)  
\end{eqnarray}

\begin{figure*}
\centerline{
\hspace{-7mm}\epsfxsize=0.45\textwidth\epsffile{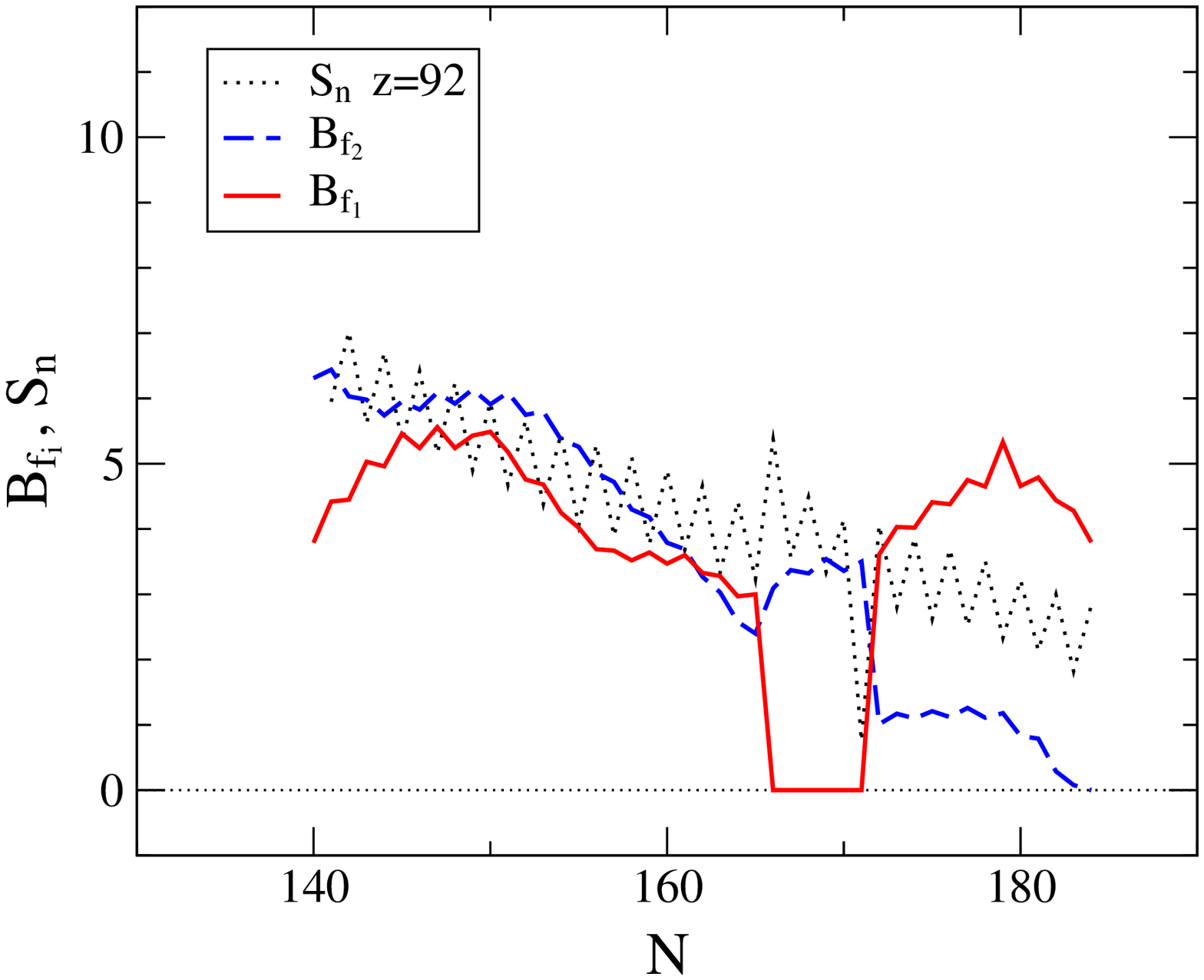}
\hspace{7mm} \epsfxsize=0.45\textwidth\epsffile{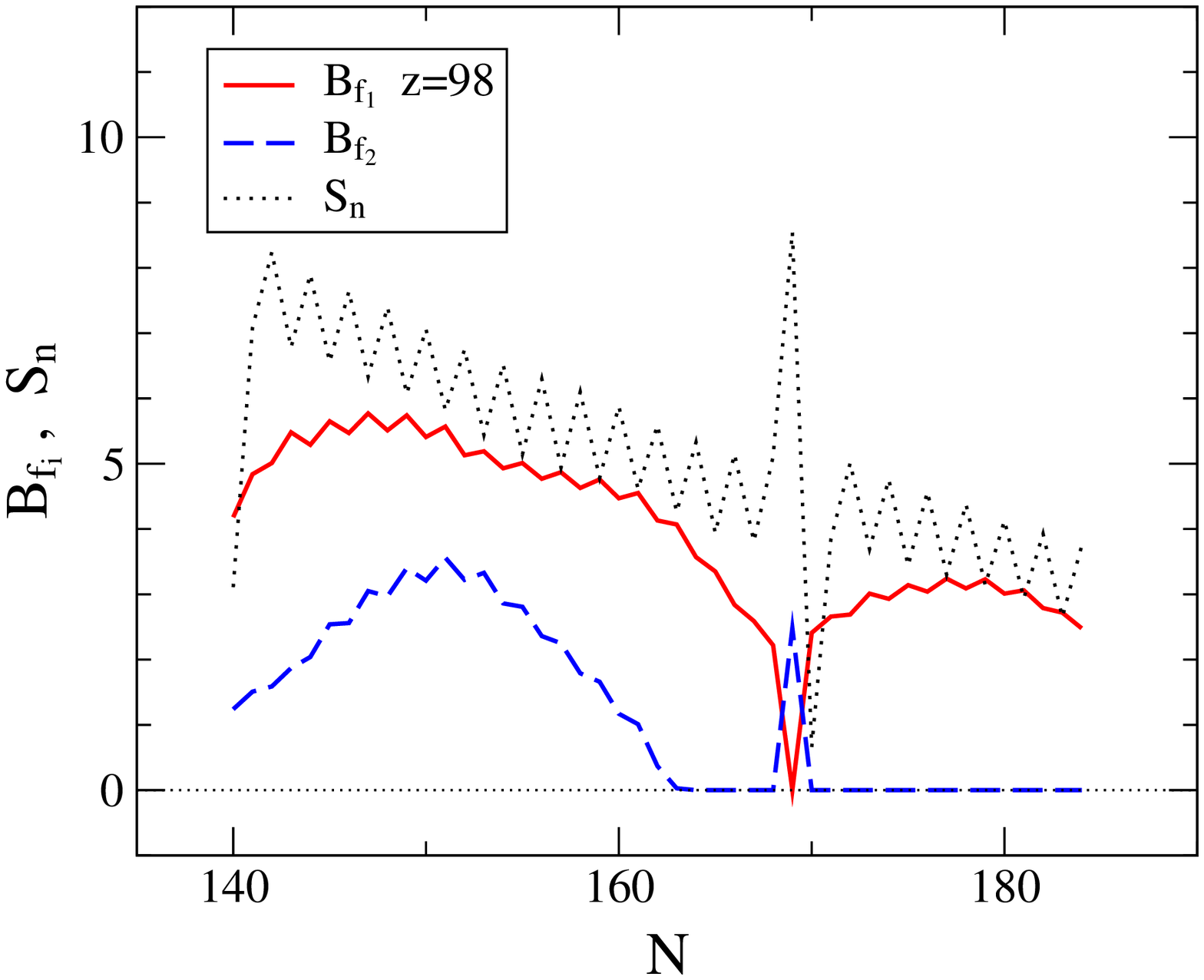}}

   \caption{
 The neutron separation energies (line) and fission barriers
 (dotted line) for U (left)  and Cf (right) isotopes on the base of Howard
 and M\"oller (1980)
 approach.
}
  \label{mssnfb}
\end{figure*}

The previous calculations \cite{TMK83,staud92} showed that
$P_{\beta dnf}$-values can  in certain cases be larger than
$P_{\beta df}$-values, especially for neutron-rich isotopes of the
Am-Cf region. In \fref{tmk83st92} some of the former results were
shown for the mass relations of Hilf et al. \cite{hilf76} and
fission barriers of ref. \cite{homo80} (\fref{mssnfb}).

In order to evaluate the role of secondary fission on the base of
new data  this process can be investigated by iterating the
SMOKER-code: if neutron emission leads to an excited level of the
residual nucleus, we again calculate the branching ratios for the
subsequent decay with the statistical model including the fission
channel. Thus the probability for fission in the second decay step
is essentially given by multiplication of the branching ratios for
neutron emission and fission of the residual nucleus. Note that
this is done by keeping track of energies, angular momenta and
parities of the excited states, and the final probability for
$(\beta^- ,nf)$-decays is obtained by summing up the contributions
from all quantum mechanically allowed partial waves.

Since the amount of excitation energy in the system is
significantly reduced with every particle emission and eventually
drops below the particle emission thresholds or fission barrier,
we found  that the cascade, could practically be terminated after
two emission steps,  contrary to neutrino-induced fission
\cite{EK} when some subsequent emissions of different particles
can occur. Thus, the residual daughter nuclei after two iterations
were assumed to be in their ground state.

The significant values of  fission probability after beta-delayed
neutron emission (second step of cascade fission) - $P_{\beta dn f
}$ can be reached only for the beta-decay with small beta-delayed
fission probability $P_{\beta df}$ (first step of the cascade) and
rather high beta-delayed neutron emission probability $P_{\beta
dn}$  of mother nucleus.   Especially high values of fission
probabilities after neutron emission (second step of the cascade)
can be achieved during decay of even-even mother nuclei (A,Z),
when neutron separation energy in final nucleus (Z+1,A-1)
significantly less than in daughter nucleus (Z+1,A) (See decay of
U with A-266, 268, 270, Table 1 in Section 4.2.2).

\section{ Survey of fission barrier  and mass predictions
 }

 As  astrophysical applications include unstable nuclei not
yet  attained by experiment, all physical quantities have to be
obtained from theoretical predictions. The quantities, which need
to be employed for the calculation of neutron-induced and
beta-delayed fission include the reaction Q-values and the fission
barrier heights. For stable nuclei, this information can come from
the experiment, but for unstable nuclei one has to make use of a
mass formula and theoretical predictions of fission barrier
heights.

 In addition to the Q-values and fission barriers involved,
 neutron emission and beta-decay require the knowledge of neutron
 optical potentials and beta strength functions.
 The accuracy of all these quantities  strongly influences the
 probabilities of beta-delayed fission \cite{{tcc89_50},{PLL90},{meyer89},{ctt91}}.
 Several predictions exist so far, but their results differ strongly due to
uncertainties for beta-strength functions and fission barriers of
very neutron-rich and especially for deformed nuclei.

\fref{tmk83st92} and \fref{mssnfb} show the connection between
fission barriers and neutron separation energies on the one hand
and fission probabilities, on the other hand, when Howard and
M\"oller predictions were used. When fission barriers are of the
order of neutron separation energy or become smaller than $S_n$,
$P_{\beta df}$ increased and reached high values.

We also considered two different up-to-date fission barrier
predictions, developed during recent years: one being based on the
Thomas-Fermi model \cite{mysw99} and another using the Extended
Thomas-Fermi model (ETFSI) \cite{mamdo01}. Both the predictions,
used in our calculations, include also the shell corrections.
     The analysis of fission barriers, predicted by
these approaches, shows that the earlier fission barriers
calculated by Howard and M\"oller \cite{homo80} are systematically
underestimated.

\begin{figure*}
\centerline{
\hspace{-7mm}\epsfxsize=0.45\textwidth\epsffile{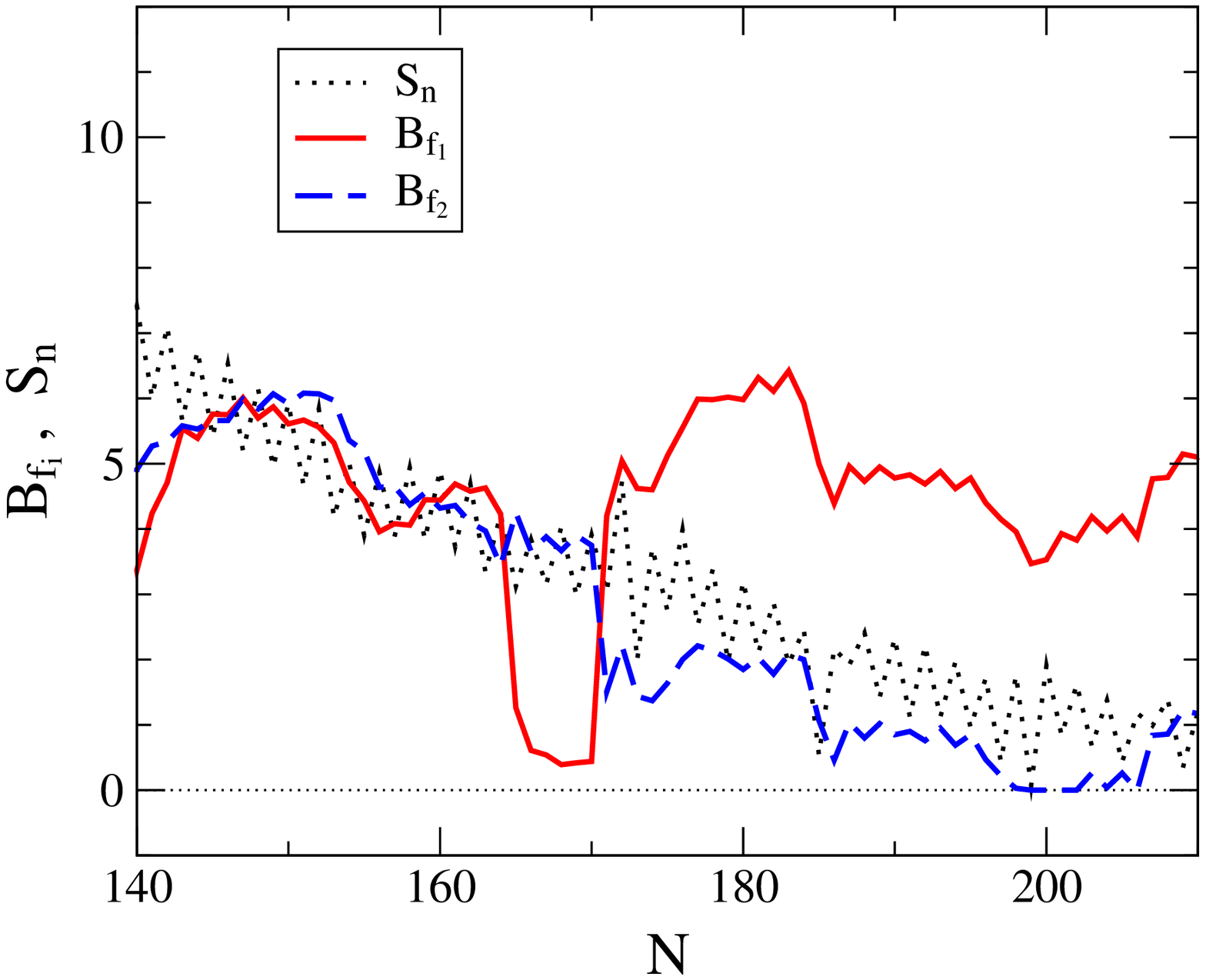}
\hspace{7mm} \epsfxsize=0.45\textwidth\epsffile{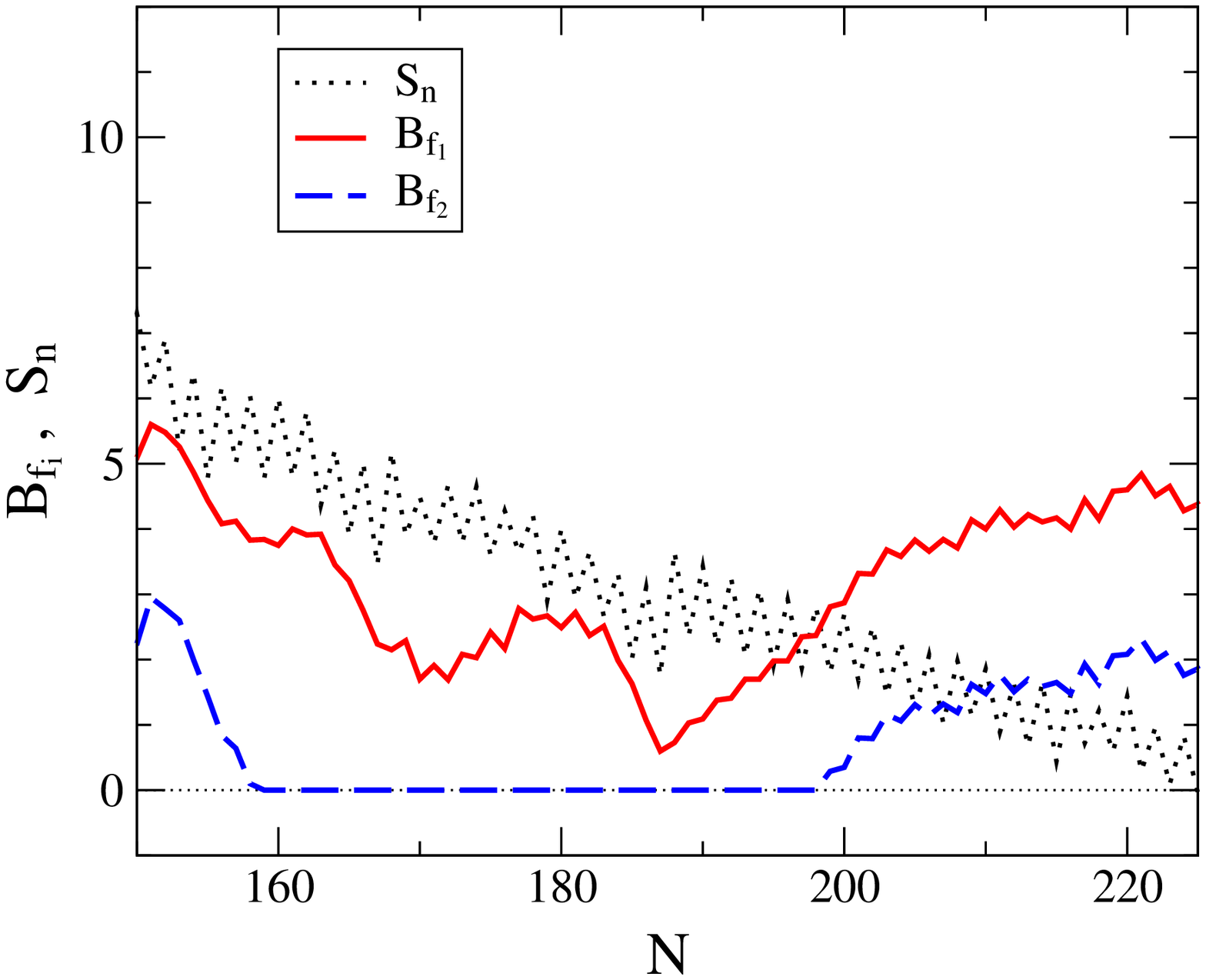}}
\vspace{8mm} \centerline{
\hspace{-7mm}\epsfxsize=0.45\textwidth\epsffile{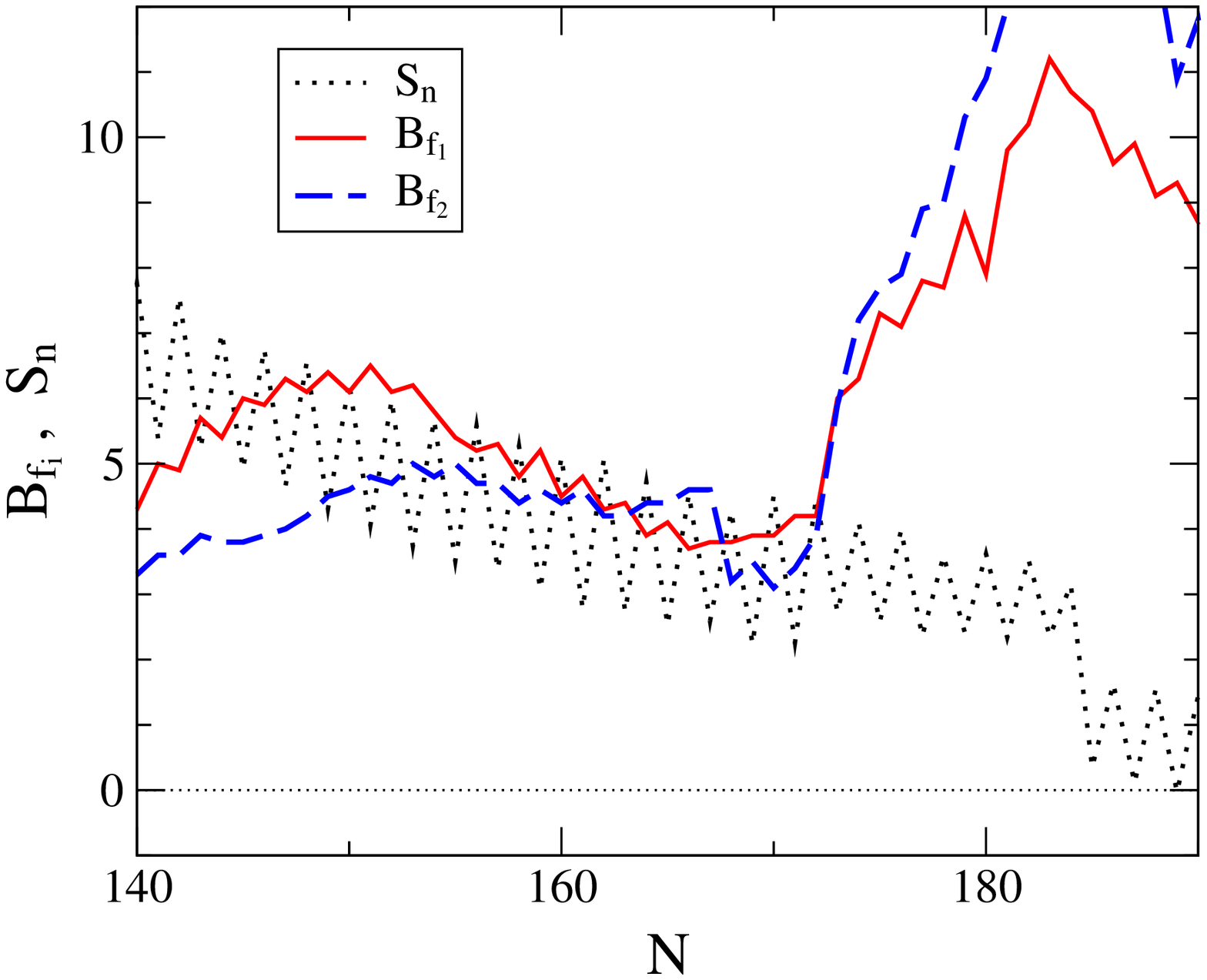}
\hspace{7mm} \epsfxsize=0.45\textwidth\epsffile{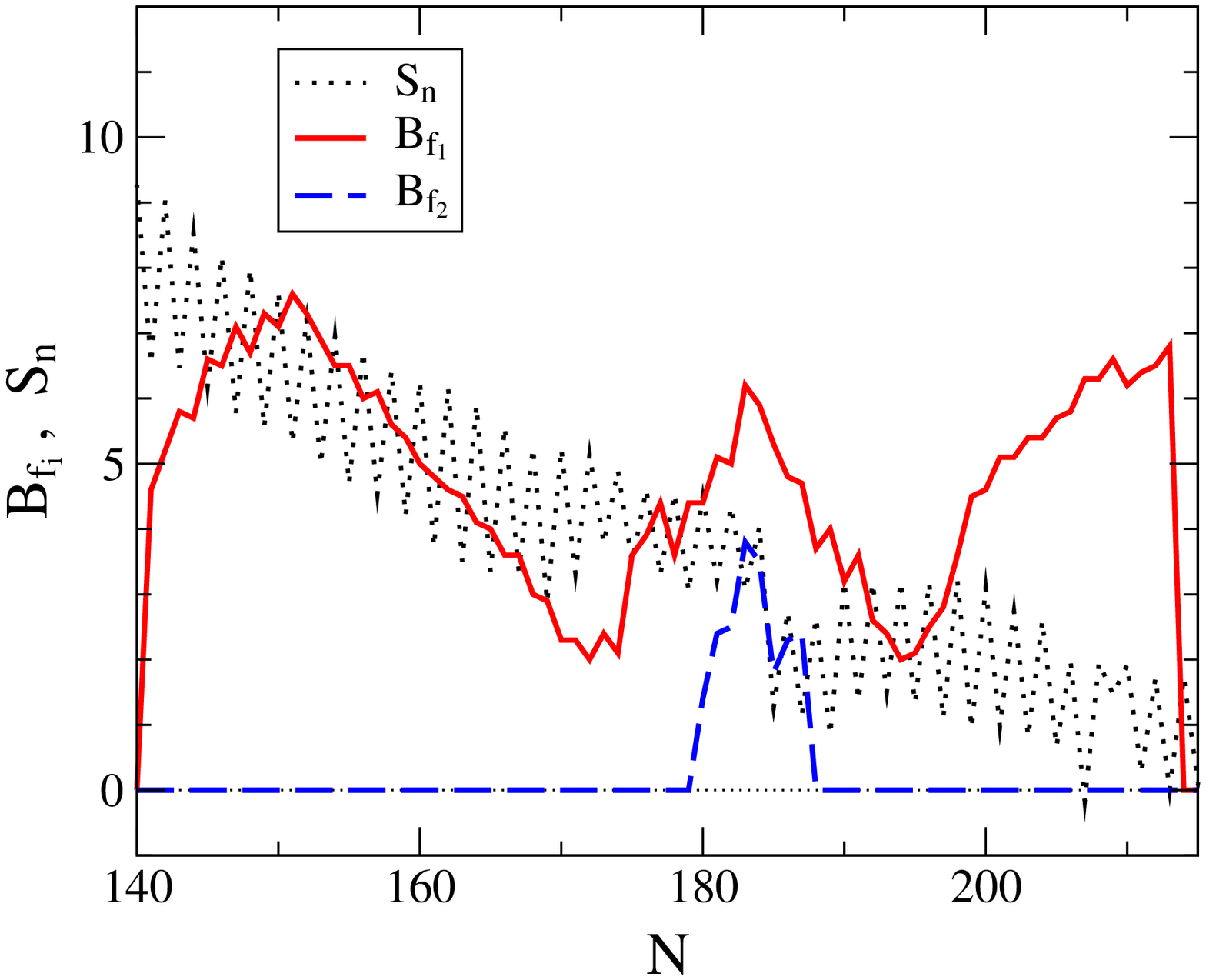}}
   \caption{
 The neutron separation energies (thin dotted line) and fission barriers
 (solid dashed line)of
 daughter nuclei under beta-decay of
 U (left) and Cf (right) isotopes on the base of different barrier predictions
 \cite{mysw99} (upper row) and \cite{mamdo01} (low row).
}
  \label{mssnfb2}
\end{figure*}

 Fission barriers  of the ETFSI model  \cite{mamdo01}
 made use of the (c,h)-parametrization \cite{brack72} (see
 detailed explanation also in \cite{bjlynn80}).
        According to the ETF-predictions, fission barrier values
increase with increasing N and that is why ETFSi predictions [13]
show significantly greater fission barrier values than the MS
calculations [2] due to several reasons, mentioned in [3,13].
  The shell corrections due to the second term of
total energy based on Strutinsky Integral can even increase the
values for nuclei with magic numbers.

 The accuracy of fission barrier predictions for
large neutron excesses  is  hard to evaluate, and its effect on
r-process results has to be studied.

The second approach \cite{mysw99} is based on self-consistent,
semiclassical mean-field solution of the problem of self-bound
nucleons interacting by suitably adjusted effective forces. This
modification of Thomas-Fermi model of nuclei has been applied to
the calculation of nuclear masses and fission barriers. To predict
fission barriers in a range of N and Z of interest a simple
algebraic expression was proposed. Predicted fission barriers
include also the shell corrections (shell-correction method is
described in \cite{mol92}). The comparison with known experimental
data shows the good agreement with calculations and one hopes the
extrapolations of fission barriers are also reliable.

 In \fref{mssnfb2}, the neutron separation energies, as well
as fission barriers for U and Cf isotopes, calculated from
considered mass relations, are shown.  As it can be seen from the
figure (the common features are the same for all actinides), the
change of barrier heights as a function of neutron number is very
similar for the different approaches, but differs strongly,
especially for N $\approx$ 184, where Mamdouh et al.
\cite{mamdo01} show  the absence of deformation and increasing
barriers in excess of  10 MeV.
 The older mass formula of Hilf et al. \cite{hilf76}, often used for
astrophysical applications, predicts significantly smaller neutron
numbers for the position of the neutron drip line than refs
\cite{homo80,mysw99}. This has important implications, because in
self-consistent calculations the r-process path  can pass nuclei
with larger neutron excess.
       For the region of interest, $A > 250$, the differences in
       $S_n$  can be larger than 1 MeV.

While the agreement for experimentally known isotopes is quite
good, the discrepancies of the approaches in the extrapolated
region can achieve some MeV.
       These  uncertainties can affect  strongly the fission rates and
the r-process. Therefore self-consistent calculations, predicting
neutron separation energies, barriers and other nuclear properties
on the same basis of the same hopefully reliable mass model are
needed.

 The preceding discussion (see also \cite{GorBf}) leads
us to the conclusion that the fission
 barrier  predictions from \cite{homo80} underestimate
whereas those from \cite{mamdo01}
 overestimate the barrier at least in the vicinity of N $\approx$ 184.
 A main problem that arises when utilizing the barriers of ref. \cite{mysw99}
 lies in the fact that only the maximum barrier height is given while
 our calculations are based on the double-hump  representation of
fission \cite{strut}.  In that case we estimated the height of the
second barrier due to barrier differences from former calculations
\cite{homo80}:
\begin{equation}
B_{f_1} = B_f([2]); \hspace{0.2cm} B_{f_2} = B_f([2])-\Delta;
\hspace{0.2cm} where \hspace{0.3cm}
\Delta=B_{f_1}([1])-B_{f_2}([1])
\end{equation}
Before  applying  the new fission rates to nucleosynthesis, we
will evaluate  the importance of individual contributions from
different fission processes, i.e.  beta-delayed fission versus
neutron-induced fission and their dependence on different mass
excess and fission barriers predictions.

\begin{figure*}
\centerline{
\hspace{-7mm}\epsfysize=0.37\textwidth\epsffile{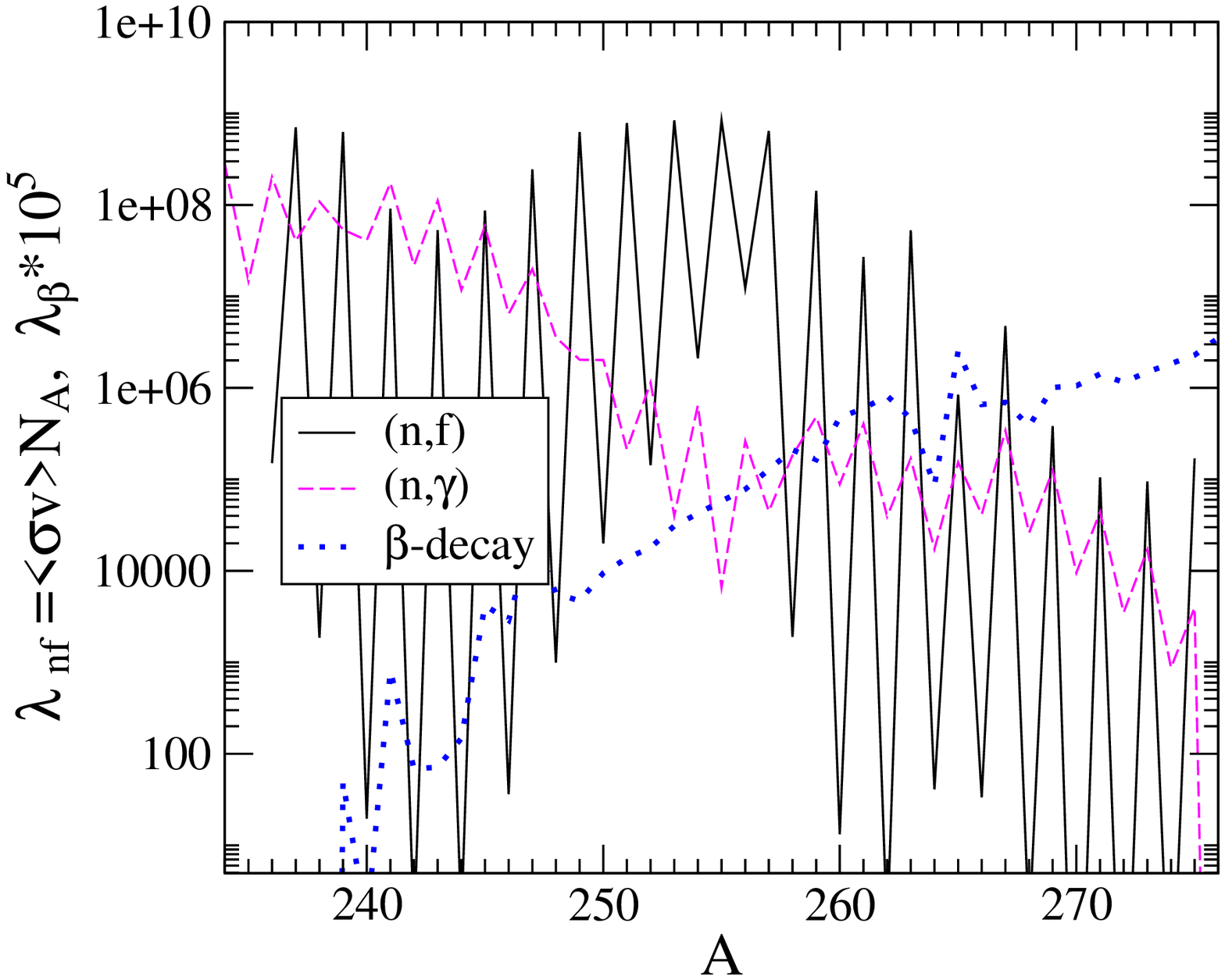}
\hspace{7mm} \epsfxsize=0.45\textwidth\epsffile{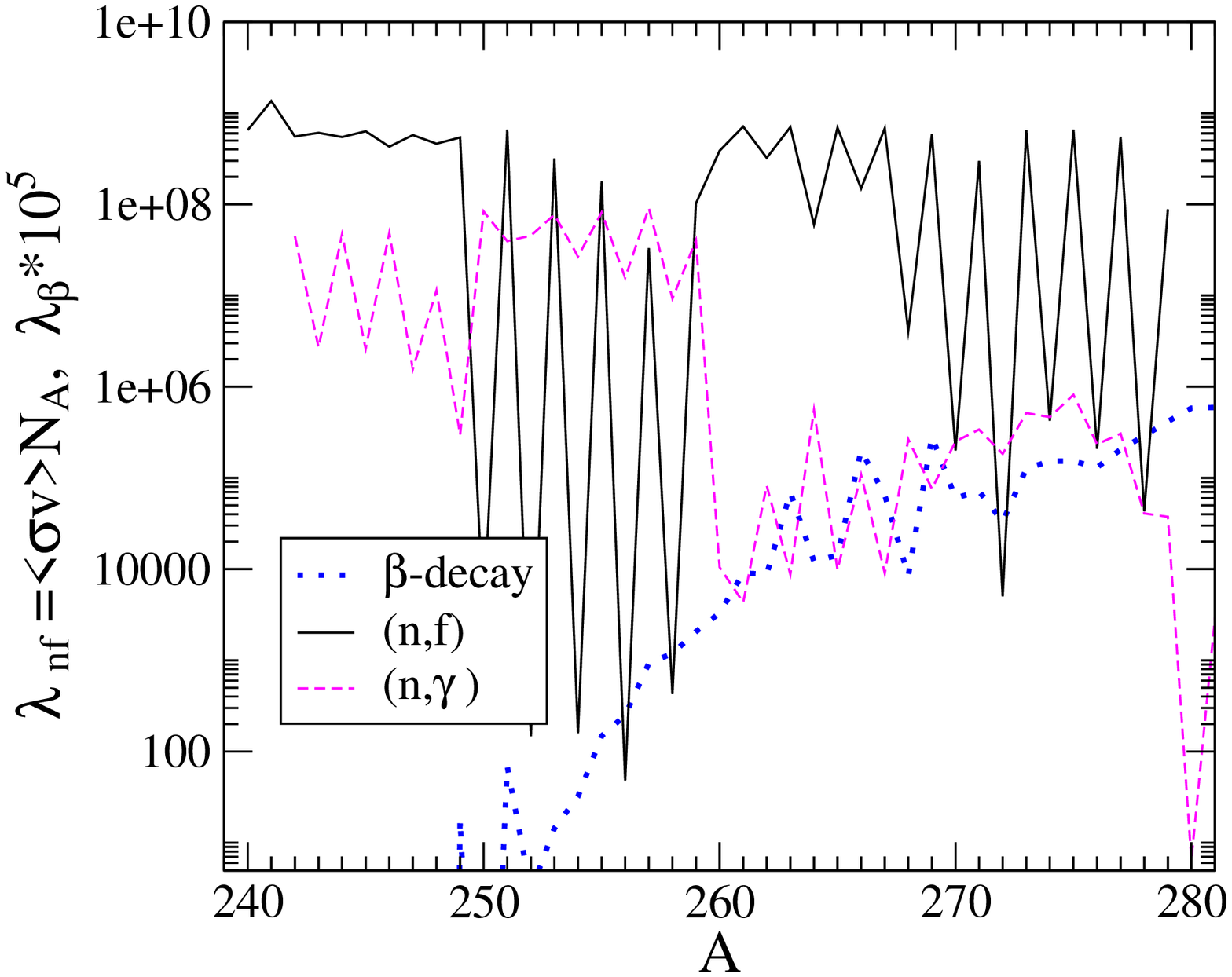}}
  \caption{The values of (n,$\gamma$), (n,f) and $\beta$-decay
  rates (in $s^{-1}$)
 for U (left) and Cf isotopes.
}
  \label{rates}
\end{figure*}

\section{Results}

The approach described above was employed for the calculations of
neutron-induced fission rates for neutron-rich isotopes up to the
neutron-drip line,  as well as beta-delayed fission rates for all
isotopes of U, Pu, Am, Cm and Cf with beta-decay energies
$Q_{\beta}>0$.

\subsection{The competition of different decay channels in
the transuranium region}

 When the r-process approaches the actinide region, fission
channels appear. Beta-delayed fission, proposed by Berlovich and
Novikov \cite{Berl69}, competes with beta-delayed neutron emission
and pure beta-decay. Also the well known process of neutron
induced fission has to be included. This process, discussed
earlier \cite{tcc89_50} for uranium isotopes has not yet been
included in r-process calculations. Neutron-induced fission rates
can be very high for isotopes of transuranium elements and exceed
beta-decay rates by orders of magnitude even near the neutron
drip-line. This is mainly the case for very high neutron densities
$n_n$ as the product $<\sigma_{nf}v>n_n$  competes with
beta-delayed fission rates $\lambda_{\beta df}$. The  recycling
timescale, measuring the time until the neutron capture on fission
 products leads again to the formation of actinides can be
  significantly less than the duration time of the r-process.
The different rates for U and Cf isotopes are compared in
\fref{rates}.
         The  rates  in  that Figure are given for $\rho Y_n=1$ and
  $T_9=1$, the beta-decay rates are multiplied by $10^5$.
 It is clear that for  $\rho Y_n > 10^{-5}$
 and for nuclei with $A>250$  neutron-induced
 fission is more important than beta-delayed fission and
 radiative neutron-capture.

 For the neutron-star merger scenario \cite{{frt99},{ross99}}
 or polar jets in supernova \cite{cam03} such conditions are maintained
 for a long time ($\sim 1 s$). However, as already pointed out above, the
  fission barriers of Howard and  M\"oller \cite{homo80} are very
  likely underestimated.

The  calculation of  rates for neutron-induced fission and
beta-delayed fission  made use of the statistical model code
SMOKER,
extended by including fission channels \cite{tcc89_50,ctt91} as
discussed in section 2.

\begin{figure*}
 \centerline{
\epsfxsize=\textwidth\epsffile{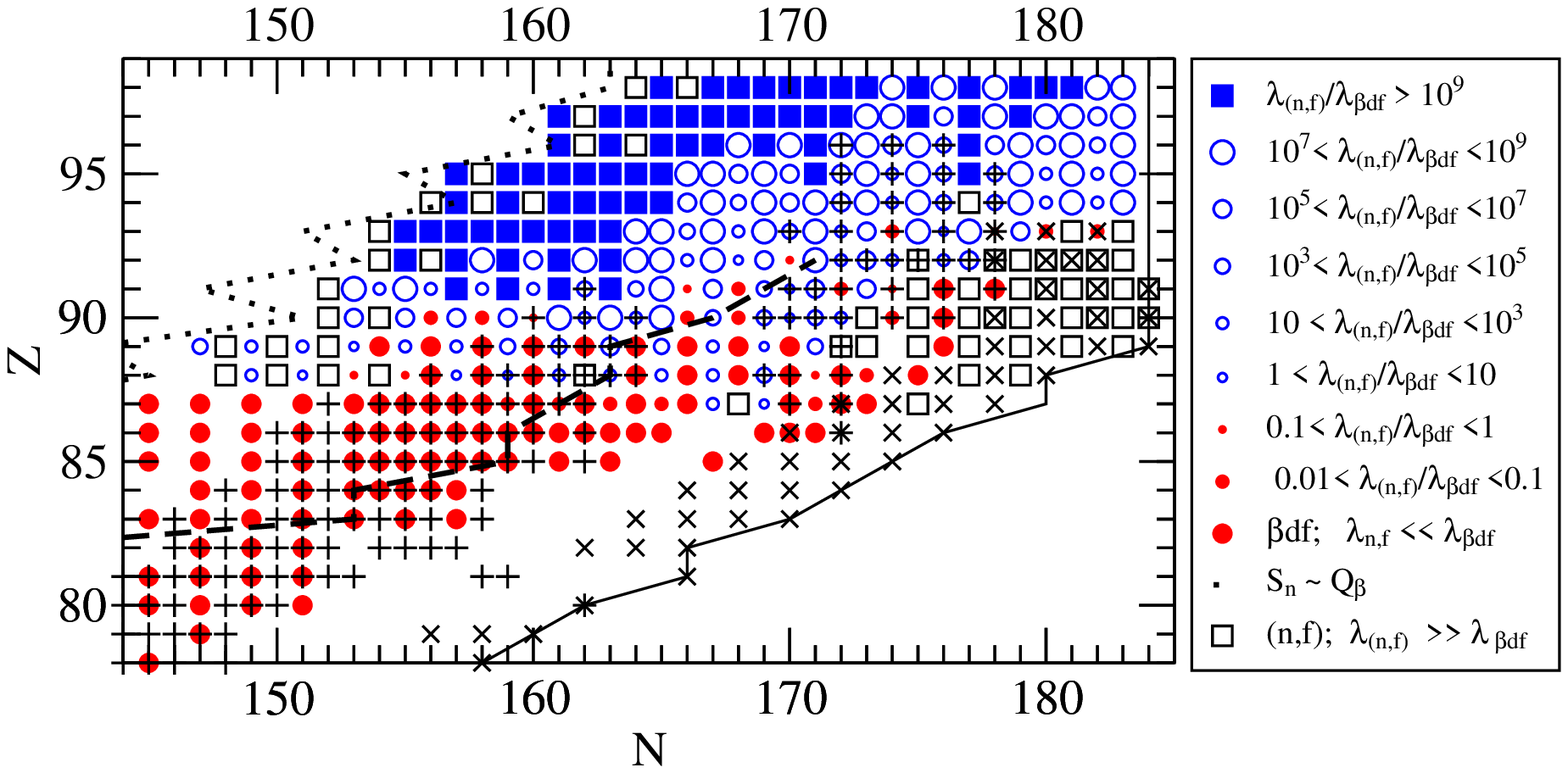}}
 \caption
  {The map of rates ratios $\lambda_{n,f} / \lambda_{\beta df}$ for $\rho Y_n=1$ and
  $T_9=1$.
}{ The most abundant nuclei along the r-process path, for
$n_n\approx 10^{26}$ (crosses), and  $n_n$  $ \approx 10^{19}$
(pluses) are marked. The position of the neutron drip-line (full
line),  nuclei with neutron separation energy $S_n \approx 2MeV$
(dashed line) and nuclei with beta-decay energies $Q_{\beta} \sim
S_n$ (dotted line) are denoted as well. These results were
obtained for the mass model of Hilf et al. \cite{hilf76} and
fission barriers by Howard and Moller \cite{homo80}.}
  \label{zamap}
\end{figure*}

A comparison of beta-delayed and neutron-induced fission rates
calculated with fission barriers of Howard-M\"oller \cite{homo80}
is presented in \fref{zamap}, where circles and squares of
different sizes indicate different ratios of rates. One can see
that for $Z\le87$ the beta-delayed fission channel is the
dominating one. For $Z=88-90$ neutron-induced and beta-delayed
fission are in competition, and for $Z>91$ neutron-induced fission
dominates. The open squares indicate nuclei for which the
beta-delayed fission rate is equal to zero. For the neutron rich
nuclei ($N>160$) this is due to the small neutron separation
energies ($ S_n<B_f$) where the decay of a daughter nucleus occurs
via neutron emission rather than fission (because the majority of
beta-decay transitions goes through low lying states). The most
abundant isotopes with Y(A,Z)$>1$\% Y(Z) (crosses) are shown for
the early stages for the r-process (when $n_n>10^{26}$) and the
time  when the neutron density decreases to less than  $10^{20} $
(pluses).

Neutron-induced fission should be of significant importance for
nuclei with the fission barrier $B_f \sim S_n$ or smaller and when
the density of free neutrons is rather high. The rates of induced
fission are large \cite{tcc89_50}, \cite{Paz03}, \cite{Jap02},
and even for higher fission barriers \cite{mysw99,mamdo01} 
 this branch of fission can be important. Preliminary calculations will be discussed
below.


As already mentioned, in the majority of the r-process
calculations fission was included in a very simple way: assuming
100\% fission at given mass number A or charge number Z in the
r-process path \cite{Chech88,frei99,rausch94}. These simplified
models permitted to investigate the behavior of nuclei formation
due to cycling but affect the accuracy of  nuclear abundances, in
part, especially cosmochronometer yields
\cite{Chech88,frei99,rausch94}.

The influence of a fission model in any astrophysical r-process
scenario becomes important when the duration time $\tau_r$ of the
r-process becomes larger than the fission cycling time
$\tau_{\mathrm{cycle}}$. In this case, the r-process can
successfully reach the transuranium region and all fission rates,
beta-delayed, and neutron-induced as well as  spontaneous fission,
should be taken into account. In case of an artificial termination
of the r-process for $A=A_{\mathrm{max}}$, the nuclei formed far
beyond uranium undergoing $\alpha$-decay will have incorrect
abundances and will lead to wrong yields of cosmochronometer
nuclei.

In some of the runs and in previous calculations \cite{pft01} we
took into account spontaneous fission only for nuclei with Z$>$92
and N$>$158 (according to known systematics) and we did not
consider neutron-induced fission for these nuclei.
 In our calculations, even
for highest fission barriers \cite{mamdo98}, the neutron-induced
fission rates are sufficiently large to terminate or at least
significantly reduce the build-up of higher Z r-process elements.

Simultaneous consideration of beta-delayed fission and
neutron-induced fission in our calculation, and spontaneous
fission in an other approach \cite{gocl99} give very similar
results (see Section 6), at least within the calculation accuracy.
However,
 neutron-induced and beta-delayed fission should be faster and
 consequently more important in the r-process  than spontaneous
 fission, which can play, however, an important role during cooling in
 the  final  decay of the r-process products.

      But for a consistent inclusion of fission in the r-process,
especially for the calculation of the final abundances, the
specific scenarios of the r-process in very high neutron density
environments should be considered.

\begin{figure*}
\vspace{10mm} \centerline{
\hspace{-1mm}\epsfxsize=0.5\textwidth\epsffile{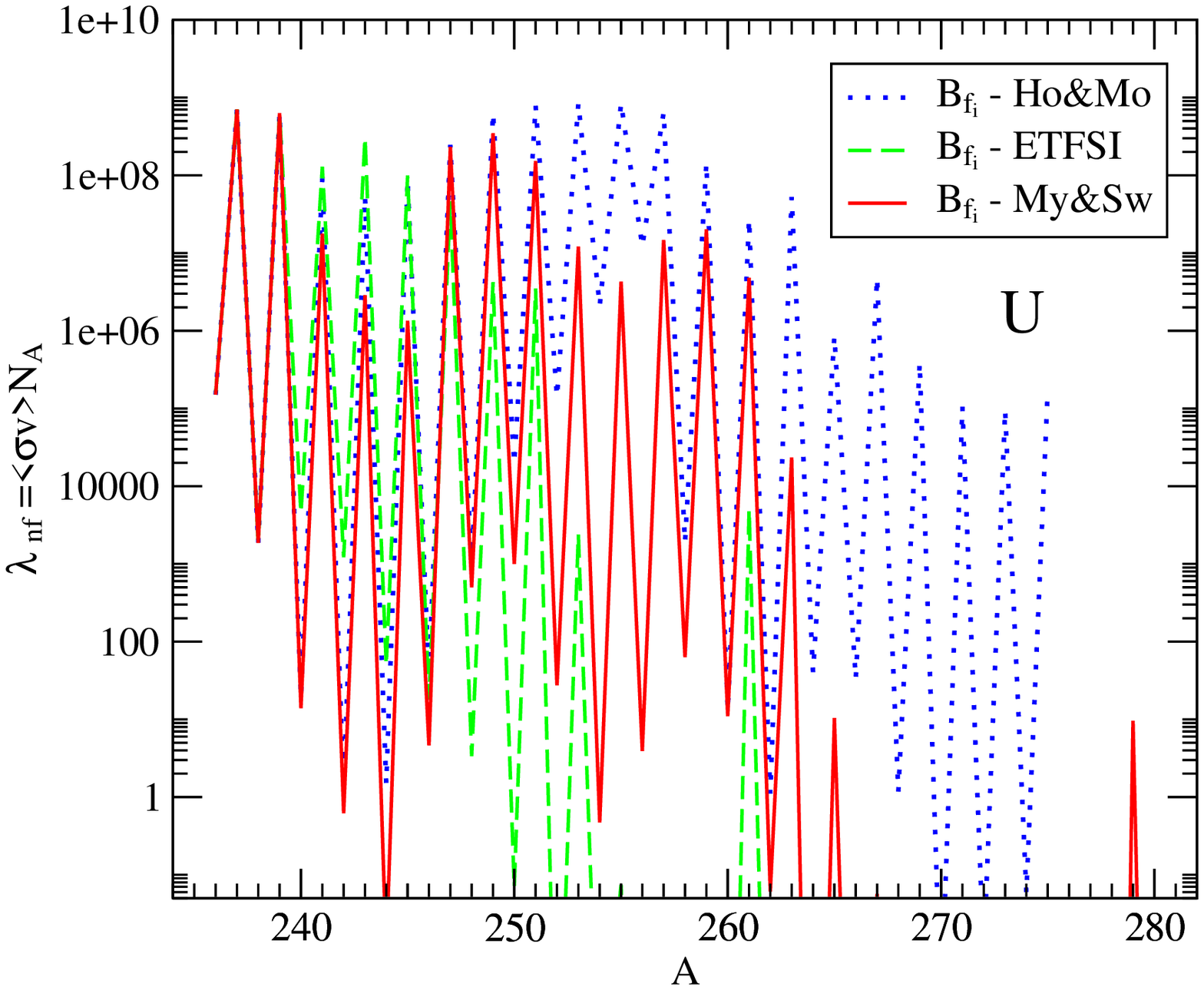}
\hspace{9mm}\epsfxsize=0.5\textwidth\epsffile{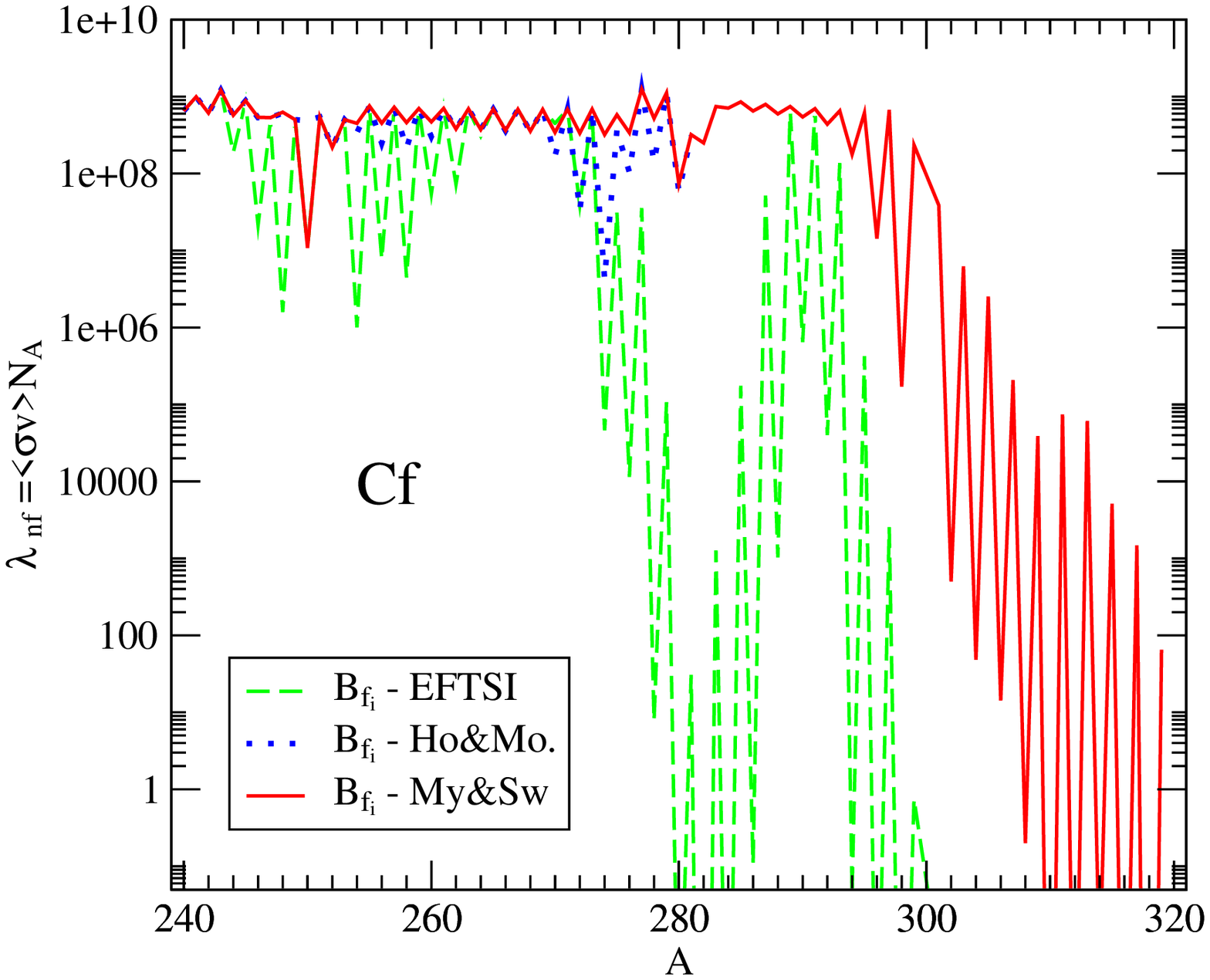}}
  \caption{Neutron-induced fission
rates for  U (left) and Cf (right) isotopes for different fission
barriers as well as for different mass formulae. }
  \label{firates}
\end{figure*}

\subsection{Mass model dependence of fission rates}

Here we present the  neutron-induced fission rates and
beta-delayed fission rates calculated  for isotopes of some
transuranium elements  taking into account different up-to-date
fission barriers and atomic mass predictions.

\subsubsection{Neutron-induced fission rates}

 In  \fref{firates}, the  neutron-induced fission rates
 $\lambda_{nf}$ derived with the barriers of ref. \cite{homo80}
and mass predictions of \cite{hilf76}    are compared with
consistent calculations when all mass  and fission barriers
predictions were made on the base of approaches  \cite{mysw96},
\cite{pear95}. The results show that for uranium isotopes with
atomic masses A$<$260 the rates using
Thomas-Fermi predictions \cite{mysw99}  are also rather high        
whereas the rates calculated on the base of ETFSI predictions
\cite{mamdo01} for atomic masses greater 250 become too small
compared to beta-decay and neutron capture rates. The latter can
not interrupt the formation of new transuranium elements in
r-process
nucleosynthesis. As mentioned above the comparison        
was made for very high neutron environments with $\rho Y_n=1$.
              As Z increases the rate values go up and
all approaches give very high values of neutron-induced fission
rates already for Cf isotopes, with the exception of the ETFSI
approach in the vicinity of neutron shell N=184 (\fref{firates}),
where $S_n << B_f$. But the heights of barriers \cite{mamdo01} for
these nuclei seem to be overestimated \cite{GorBf}.


\subsubsection{beta-delayed fission rates}

\begin{figure*}
\centerline{
\hspace{-2mm}\epsfxsize=0.5\textwidth\epsffile{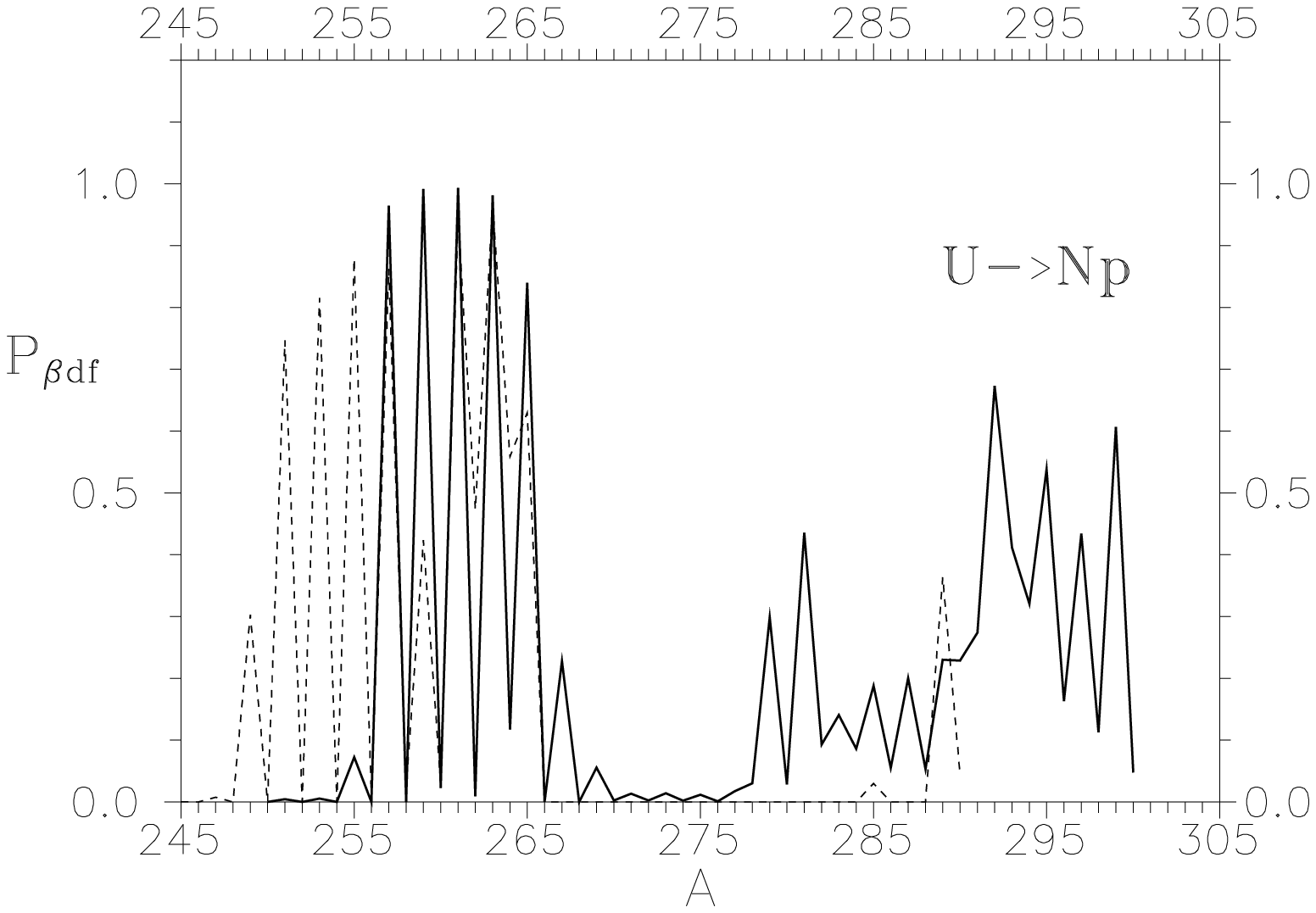}
\hspace{2mm}\epsfxsize=0.5\textwidth\epsffile{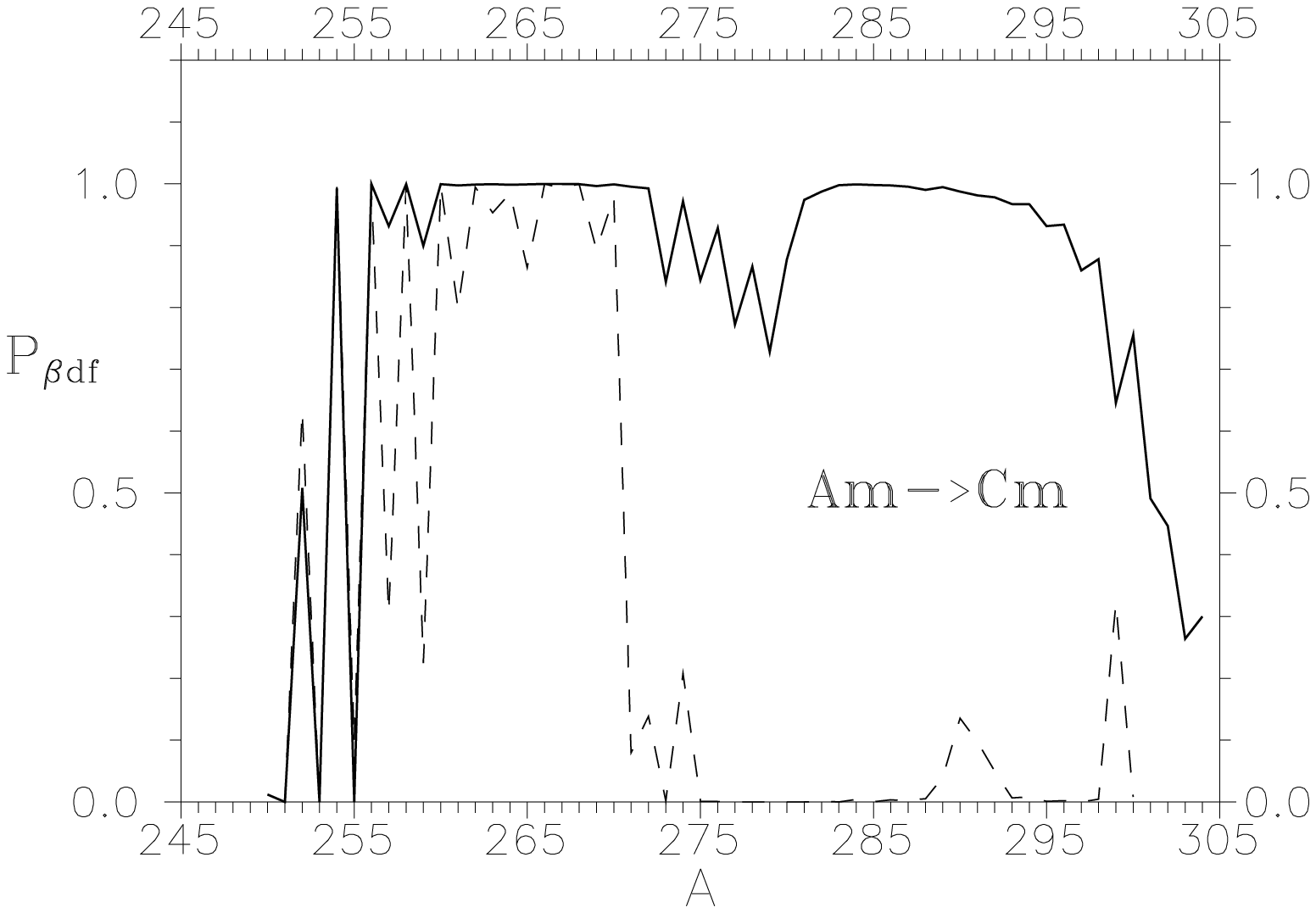}}
\centerline{
\hspace{-2mm}\epsfxsize=0.5\textwidth\epsffile{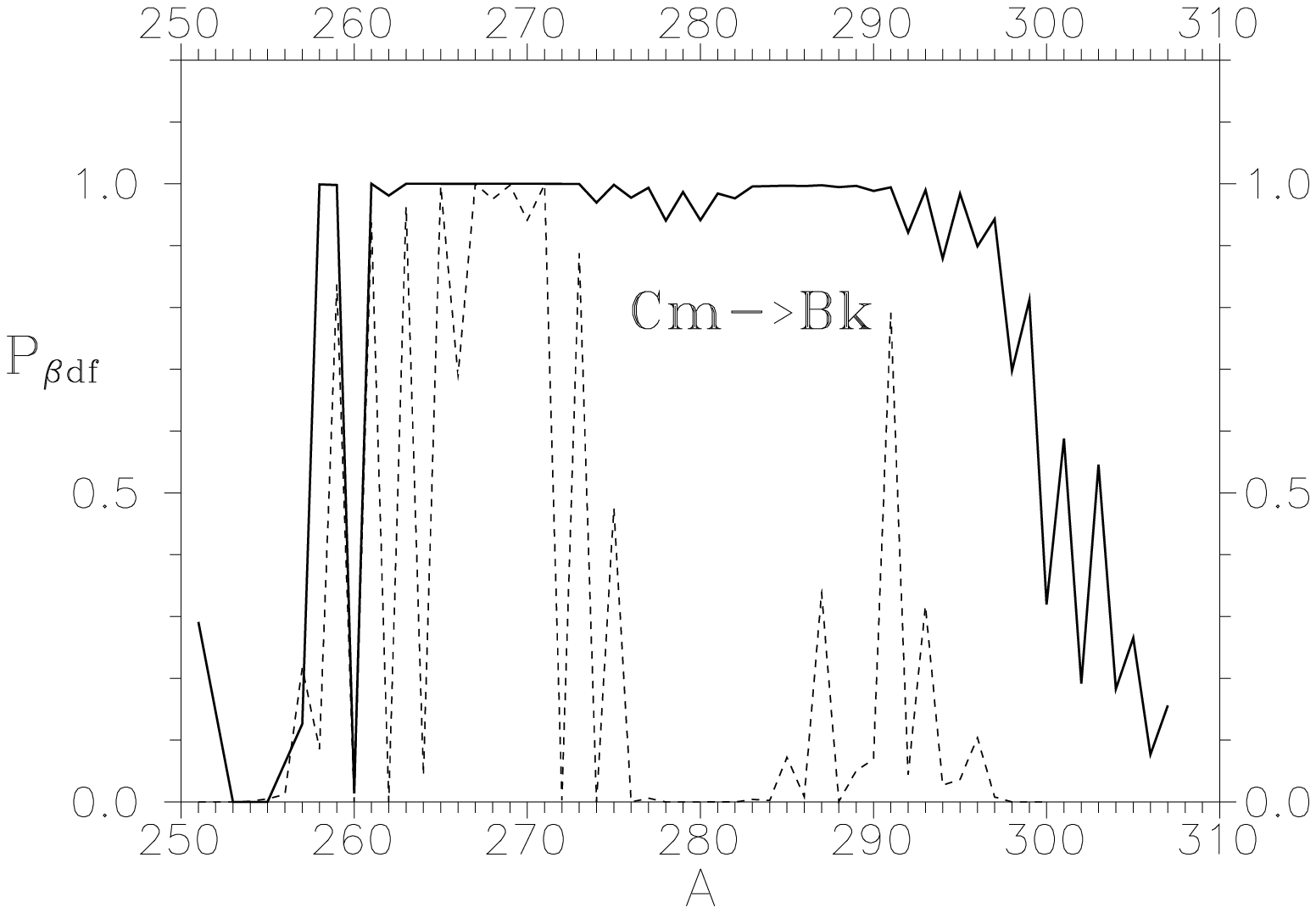}
\hspace{2mm}\epsfxsize=0.5\textwidth\epsffile{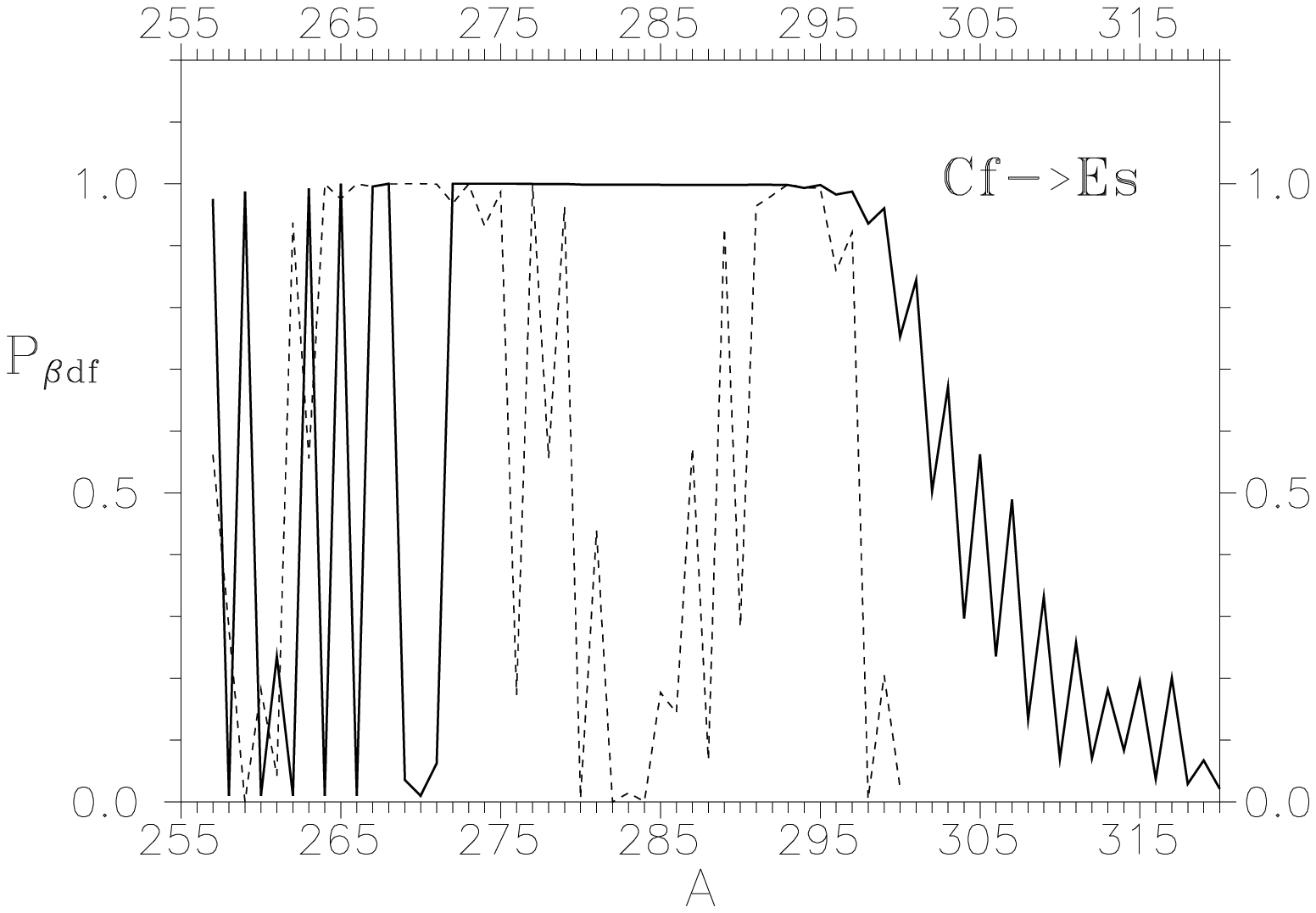}}
\vspace{2mm}
  \caption{Fission probabilities $P_{\beta df}$
 for  U, Am, Cm and Cf  isotopes for  mass
 models of Mamdouh et al. (Dashed line) and Myers and Swiatecki (line). }
  \label{pbdf_ms}
\end{figure*}

The first extended results for beta-delayed fission probabilities
(\fref{pbdf_ms}) for the majority of neutron-rich isotopes of some
chemical elements show that  for fission barriers of refs.
\cite{mysw99}, \cite{mamdo01},  which are significantly higher
then previously used \cite{homo80}, the values of $P_{\beta df}$
are still high for very neutron rich isotopes. For majority of
them this applies in the case when mass model \cite{mysw99} was
used and for part of them  when the  model based on ETFSI
\cite{mamdo98} was used.

 Calculations of fission
rates in the framework of the  statistical model(see \cite{TMK83}
and Section 2 of the present work) for both self-consistent
approaches \cite{mamdo01}, \cite{mysw99} show similar results when
the maximum values are reached for isotopes with atomic mass
numbers A$\approx 255-265$ .
        But  with increasing Z the region of isotopes with high values of
$P_{\beta df}$ close to 100\% expands and in case of data set
\cite{mysw99} the majority of isotopes attain close to maximum
values above  Z=95 and in case of \cite{mamdo01} above Z=98. The
suppression in $P_{\beta df}$ values for californium for the ETFSI
barriers is explained by very high fission barriers in the
vicinity of nuclei with number of neutrons close to 184.

In Table 1, the results of the cascade calculations for
beta-delayed fission are shown. We present the results of the
calculations for the nuclei with very small values of direct
beta-delayed fission probability $P_{\beta df}$. As it was
expected the fission probability after delayed neutron emission
$P_{\beta dnf}$ can be high. For some decays total
$P^{\mathrm{tot}}_{\beta df}$ which is the sum of beta-delayed
fission (first step of cascade fission) and fission after
beta-delayed neutron emission (second step of cascade) increases
by several times. That means, that the cascade mechanism can be
important for nuclei for which the direct (one step) beta-delayed
fission is small and beta-delayed neutron emission is high.

\begin{center}
\begin{table}
\caption{Beta-delayed cascade fission probabilities (in \%) for
some U  isotopes.$^{1)}$}
\begin{tabular}{|c|c|ccc|c|cc|}
\hline
 decay &A& $P_{\beta df} $   & $P_{\beta n}$    &$P_{\beta dn f}$ &
 ${P_{\beta df}^{\mathrm{tot}}}$& $P_{n\beta df}^{\mathrm{max}}$ & $P_{n\beta
 df}^{\mathrm{min}}$
  \\[1mm]\hline \hline
  U $\rightarrow$Np & 264 & 41  &  59  &   4   &  45  &   86& - \\
 U $\rightarrow$Np & 266 & 8  &  92 &   69   &  77  &   38& - \\
U $\rightarrow$Np & 268 & 1  &  99  &   28   &  29  &   25& 7 \\
U $\rightarrow$Np  &270    & 3  &  96  &   23 &   26  &   44 & 14\\
U $ \rightarrow$Np & 272   & 3  &  97  &   8  &   11   &   23 & 10 \\
U $\rightarrow$Np  & 274   & 4  &  96  &   4  & 8  &   16 & 11 \\
U $\rightarrow$Np  & 276   & 2  &  98  &   9  & 10  &   31 & -
\\[1mm]
\hline\hline
\end{tabular} \\
$^{1)}$   $f{P_{\beta df}^{\mathrm{tot}}}$(Z,A)=$P_{\beta df}+P_{
\beta dnf}$,   \hspace{5mm} $P_{\beta dnf}^{\mathrm{max}} \equiv
P_{\beta df}(Z+1,A-1) $,   \hspace{5mm}
\end{table}
\end{center}

In the last two columns the results of our  evaluations are
presented. The first ($P_{\beta dnf}^{\mathrm{max}}$) shows the
fission contribution during beta-decay of daughter nucleus. The
second one ($P_{\beta dnf}^{\mathrm{min}}$) shoes the same but in
assumption that compound nucleus does not changed during all the
steps of cascade.
        It was calculated as probability of delayed fission of
        daughter nucleus but with reduced $\beta$-strength function
        of the mother nucleus $S_{\beta}(U)$ where $U=E-S_n$ as in \cite{lps85}.
Different mass and barrier predictions for U \cite{homo80} and Cf
\cite{mamdo01} were used.

\section{Discussion and outlook. }

\subsection{
 Consequences of
new rates utilization for the r-process) }

The discrepancy of fission rates due to different mass predictions
is high, especially for the region of nuclei with N $\sim$ 184,
where different mass models predict different dependence of
fission barriers on N. New efforts are needed to improve the
models for better agreement, mainly in the region of nuclei with
very big neutron excess so important for the r-process. It should
be noted that the results of Hartree-Fock mass calculations
\cite{GorBf} point out that ETFSI-calculation \cite{mamdo01} are
overestimated, at least in the vicinity of N=184. That is why we
can consider the fission rates of Howard and M\"oller
\cite{homo80} were underestimated, the rates based on ETFSI  are
overestimated. This assumption lets us discuss what can be
expected when the fission rates are applied to the r-process
calculations.

\begin{figure*}
\vspace{10mm} \centerline{
\hspace{-7mm}\epsfxsize=\textwidth\epsffile{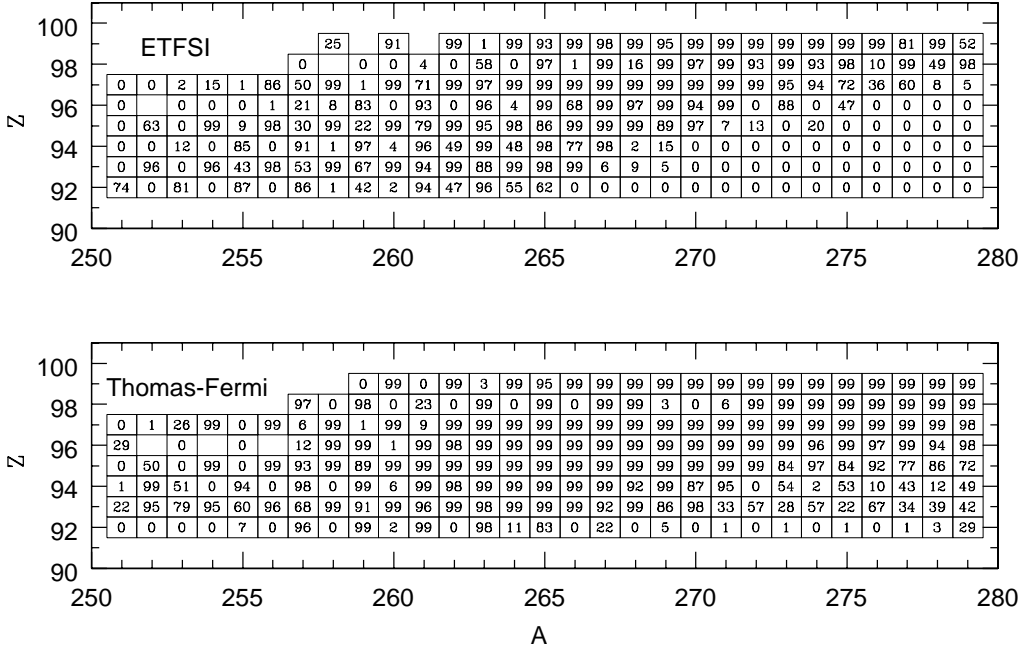}}
  \caption{ETFSI and Thomas-Fermi beta-delayed fission probabilities }
  \label{bdfrates}
\end{figure*}

Looking at \fref{bdfrates} and Table 2, it is clear that the
beta-delayed fission probabilities are quite high for the majority
of isotopes of transuranium elements, independent of differences
in mass models and strength-function calculation.

When moderate neutron densities are considered ($n_n \sim 10^{22}
- 10^{24}$ for temperature of 1.3$\times 10^9$ K), the r-process
passes through the region of nuclei with neutron separation
energies $S_n \approx 2-3 MeV$. Very high fission barriers, as
predicted by some calculations \cite{mamdo01} will not avoid an
influence on the r-process, they only move the effective region to
nuclei with atomic mass numbers A $\approx$ 260, where fission
rates, particularly beta-delayed fission, are high
(\fref{pbdf_ms}, \fref{bdfrates}). For very high neutron densities
($n_n
>> 10^{24}$), when the path of the r-process lies closer to
neutron drip line,  beta-delayed fission rates for fission
barriers by Mamdouh et al. \cite{mamdo01} will be small. But
neutron-induced fission rates can be significant for such
conditions and additional calculations of the r-process are needed
to clarify the question.

\begin{center}
\begin{table}
\caption{Beta-delayed fission probabilities for U and Cf
isotopes.}
\begin{tabular}{|c|cccc||c|cccc|}
\hline
%
 U & $HoMo^{*1)}$ & $HoMo$ &$MsSw$ & ETFSI& Cf & $HoMo^{*1)}$ &
  $HoMo$ &$MsSw$ &ETFSI
  \\[1mm]\hline \hline
  250 &  0  &   0  &   0  &   0  &    260  &   0   &   0  &   01 &  0\\
  251 & 86  &  86  &   0  &  74  &    261  &   0   &   4  &   23 &  4\\
  252 & 22  &  33  &   0  &  00  &    262  &   0   &   0  &   01 &  0\\
  253 & 94  &  94  &  01  &  81  &    263  &   08  &   71 &   99 &  58\\
  254 & 71  &  76  &   0  &   0  &    264  &   01  &   05 &   01 &  00\\
  255 & 98  &  99  &  08  &  88  &    265  &   80  &   99 &   99 &  97\\
  256 & 91  &  94  &   0  &   0  &    266  &   83  &   87 &   01 &  01\\
  257 & 99  &  99  &  96  &  88  &    267  &   95  &   99 &   99 &  99\\
  258 & 89  &  96  &  00  &  03  &    268  &   90  &   99 &   99 &  16\\
  259 & 98  &  98  &  99  &  75  &    269  &   91  &   99 &   04 &  99\\
  260 & 40  &  93  &  03  &  01  &    270  &   81  &   99 &   01 &  97\\
  261 & 95  &  99  &  99  &  96  &    271  &   88  &   99 &   06 &  99\\
  262 & 15  &  88  &  01  &  12  &    272  &   72  &   99 &   99 &  93\\
  263 & 86  &  97  &  96  &  98  &    273  &   85  &   99 &   99 &  99\\
  264 & 38  &   6  &  12  &  73  &    274  &   77  &   97 &   99 &  93\\
  265 & 37  &  97  &  83  &  76  &    275  &   88  &   99 &   99 &  98\\
  266 & 26  &  91  &  00  &  00  &    276  &   69  &   80 &   99 &  10\\
  267 & 24  &  97  &  20  &  01  &    277  &   81  &   99 &   99 &  99\\
  268 & 01  &  27  &  00  &   0  &    278  &   81  &   99 &   99 &  49\\
  269 & 44  &  94  &  05  &   0  &    279  &   86  &   99 &   99 &  98\\
  270 & 05  &  32  &  00  &   0  &    280  &   88  &   99 &   99 &  01\\
  271 & 23  &  85  &  01  &   0  &    281  &   86  &   99 &   99 &  84\\
  272 & 05  &  35  &  00  &   0  &    282  &   87  &   99 &   99 &  00\\
  273 & 16  &  60  &  01  &   0  &    283  &   89  &   99 &   99 &  05\\
  274 & 03  &  38  &  00  &   0  &    284  &    -  &    - &   99 &  00\\
  275 & 30  &  74  &  01  &   0  &    285  &    -  &    - &   99 &  98\\
  276 & 07  &  37  &  00  &   0  &    286  &    -  &    - &   99 &  99\\
  277 & 46  &  97  &  02  &   0  &    287  &    -  &    - &   99 &  99\\
  278 &  -  &   -  &  03  &   0  &    288  &    -  &    - &   99 &  20
\\[1mm]
\hline\hline
\end{tabular}    
$^{1)}$ Mass predictions were made according to ref. \cite{hilf76}
\end{table}
\end{center}

In every case of long neutron exposure (with durations of the
r-process $\tau_r >> \tau_{\mathrm{cycle}}$), the  r-process seeds
capture on average more than 150 neutrons, thus moving high
abundances to the end of r-process path, where the fission stops
the formation of heavier nuclei. Due to fission, the fission
products enter the r-process again at smaller mass numbers
(r-process cycling). The termination of the r-process due to the
fission occurs in the range of nuclei with charge numbers
$92<Z<98$. Here, $\tau_{\mathrm{cycle}}$ - the time of r-process
propagation from the average fission products to the transuranium
region is defined as :
$$
\tau_{\mathrm{cycle}}^{-1} = \sum_{Z_i}\lambda_{\beta}^i;
\hspace{0.3cm} \lambda_{\beta}^i=\sum_j Y(Z_i,
A_j)\cdot\lambda_{\beta}^j
$$
For higher fission barriers a longer duration time is encountered
and
 heavier nuclei are formed.

\subsection{Possible role of
triple/ternary fission in the r-process}

Utilization of the latest calculations of fission barriers
\cite{{mysw99},{mamdo01}} should result in a decrease of fission
rates in comparison to former barriers \cite{homo80} previously
used in r-process calculations. This in turn should lead to an
increase of the maximum atomic numbers of fissioning nuclei formed
in the r-process in excess of Z$\approx$96.

In comparison to the Hilf et al. mass formulae \cite{hilf76},
recent consistent approaches shift the neutron drip line in the
direction of heavier masses. For example, the mass prediction of
Hilf et al. \cite{hilf76} defines the atomic mass of the heaviest
Cf isotope to be A=279, while the other calculations discussed
above \cite{{mysw99},{mamdo01}} predict the existence of Cf
isotopes heavier than A=300.
         The widening of the r-process region can result in a shift of
the r-process path into the direction of larger atomic mass
numbers. In case of r-process environments with moderate  neutron
densities binary fission leads  to an abundance curve downing
strongly below $A \sim 130$ due to cycling of the r-process
 after all lighter nuclei have vanished due to neutron
 capture.  For very high neutron density environments, when the
r-process passes partly along or very close the neutron drip-line,
the mass of light fission fragment can be greater than 130. Such a
mass distribution of fission fragments could affect the good
agreement of calculations and observation of heavy element
abundances when using binary fission only.

 Triple or ternary
fission \cite{{ternmas},{ternPere},{tripl}} can populate nuclei
with $A<130$ and should be considered as well. It can save the
agreement with observations, but to define the exact contribution
of triple fission into the r-process the extended nucleosynthesis
calculations should be done.

Ternary and quaternary fission are observed  experimentally
\cite{open3_4}. Theoretical evaluations of the possibility of
fission into {\it n} equal fragments (ternary fission when $n=3$)
followed \cite{theor34}. For example, when the fissility parameter
exceeds 0.61 (that is usual for neutron rich transuranium nuclei),
ternary fission energetically becomes more preferable  to binary
fission.

       Of course, the probability of triple fission of long lived
or experimentally known nuclei is small because of the high second
barrier of ternary fission and long path to the saddle  point of
ternary fission in comparison with binary fission.
       The experimental evaluation  of the ternary fission probability in neutron
induced reactions at low energy shows that this process is at
least by 3 orders smaller than binary fission and the formation
probability  of a heavy third fragment (A$>40$) decreases below
$10^{-6}$ \cite{tripl}.
      The latest extended
experiment using the FOBOS-detector \cite{ternPenj} shows that
fission yields for three fragments with $Z>20$ are approximately
$10^{-3}$ lower than the binary ones in the nuclear system formed
in the $^{14}N+^{234}Th$ collision. In forthcoming experiments the
ternary yields in the decay of nuclei with larger $Z^2/A$ will be
measured \cite{ternPenj}.
        But
 experiments with heavy ions \cite{ternPere} show that the yield
of ternary fission fragments of stable or long-lived isotopes of
Th, U and Bi  can be of the same order as in binary fission. The
masses of ternary fragments were approximately equal.

 Up to now there are no theoretical
 predictions for ternary fission of unstable neutron-rich nuclei involved
 in the r-process \cite{ternmas}. According to the theory of ternary
fission, fission into three fragments occurs if reach the
quasi-stable ternary valley \cite{ternmas}, which is difficult to
investigate experimentally  because of the high second barrier.
But with increasing atomic number and probably neutron number, the
second barrier of the ternary valley decreases, especially when
the lightest third fragment $A_3$ is significantly heavier than
alpha-particles, and the probability of ternary fission increases.

All these experimental facts and theoretical evaluations let us
consider that after beta-delayed fission of extremely neutron rich
nuclei (when the energy of beta-decay is about 15 MeV or  more and
is of the order of shell corrections), the probability of ternary
fission of nuclei with fissility $x>0.61$ can be rather high.

In the extreme case when ternary fission of very neutron rich
nuclei is strong, many intermediate nuclei can be formed in the
r-process  and the impact  of such a fission process upon the
yields of intermediate r-elements should be investigated.

At least two extreme cases can be discussed:

1) true ternary fission into 3 approximately equal fragments
  (quatro fission occurs when $x>0.87$ and it is not considered, because  the fissility
  values for neutron-rich nuclei involved in r-process lie
  mainly  in the x=0.55-0.7 region).

2) triple fission when the average mass of the third fragment is
small: $A_3\sim10-20$. In this case two other fragments should
have smaller masses than in the case of simple binary fission
because 10-20 mass units can be gathered in the third fragment. At
least one heavy fragment will be included into the r-process
before the $A=130$ peak.

\section{
 Summary}


The new calculations of fission rates considered two consistent
approaches. The calculations showed that, on a whole, the new
rates based on fission barriers of \cite{{mysw99},{mamdo01}} are
lower than the ones previously used
\cite{{TMK83},{Paz03},{Jap02}}. However, taking into account a
cascade model of  fission after beta-delayed neutron emission, the
total value of fission rates can increase in some cases,
especially when beta-delayed fission of the mother nucleus is low
and primary beta-delayed neutron emission is high. For such nuclei
the addition to delayed fission after neutron emission can be
significant and the total value of beta-delayed fission can
increase strongly.

Unfortunately, the agreement of different calculations (used in
different mass and fission barrier predictions) have not converged
yet, especially for very neutron rich nuclei in the region with N
$\sim $
 184.

 The extended calculations of fission rates taking into account
 Thomas-Fermi \cite{mysw99} and ETFSI \cite{mamdo01} fission barriers
 have not been finished yet.
 A detailed
comparative analysis of the influence of fission modes on the
results of nucleosynthesis was made in full network calculations
\cite{{Jap02},{Paz04}} using the mass relations \cite{hilf76} and
fission barriers \cite{homo80}, used for years in many
astrophysical applications.
        An evaluation of the relative
contribution of neutron-induced and $\beta$-delayed fission on the
formation of r-elements under  conditions of high neutron
densities ($10^{20}<n_n<10^{30}$) was made. Our  present
calculations of
 fission rates should not change significantly our previous
 conclusions concerning competition of beta-delayed fission
 and neutron induced fission but the resulting changes in
  r-process abundances can be significant.

The calculation of the yields Y(A) for the range of atomic masses
$120<A<240$ depend weakly on the model of the fission
approximation. For predicting the yields of nuclei heavier than
240, especially nucleo-cosmochronometer yields and abundances of
intermediate nuclei with atomic mass numbers 100$<A<$120, fission
should be included as precisely as possible.
      Nuclei lighter than 120 can be formed  if
the fission of many transuranium nuclei is mainly asymmetric. To
consider all the opportunities, a schematic model for the mass
distribution \cite{pft01}, including both symmetric and asymmetric
\cite{Smir89} fission was applied earlier.

In the classical model of the r-process, the r-process path
proceeds under moderate free neutron densities  and close to
nuclei with a neutron separation energy $S_n \approx$ 2MeV, for
which in the transuranium region the values of beta-delayed
fission are the highest ones. According to the existing fission
probability calculations $P_{\beta df}$ \cite{TMK83,staud92},
based on old fission barrier predictions \cite{homo80}, the
r-process would stop in the  region of transuranium nuclei at
atomic mass numbers A~$\approx~$250-260 and charge Z
$\approx$90-96. In that case,
 the approximation of 100\% instantaneous fission in
the vicinity of A$\sim$260 (for simplicity A=260) used in
\cite{{cowan99},{rausch94},{tcc89_50}} would be quite good.
However, new fission rates can change the termination point of the
r-process to the region of A~$\approx~$260-280 and Z
$\approx$94-98.

When considering dynamical r-process models at extremely high
neutron densities, in which fission occurs strongly, the path of
the r-process lies close to neutron drip-line, and the efficiency
of beta-delayed fission depends strongly  on fission rates for the
group of nuclei close to the neutron shell N$\approx$184. In
particular, in the scenario realized in the model of neutron star
mergers  \cite{frt99}, \cite{ross99},  r-process nucleosynthesis
proceeds mainly along the neutron drip-line \cite{pan_path} due to
very high neutron densities. In this case, the role of
neutron-induced fission will be comparatively more important than
beta-delayed fission.
  For realistic fission barriers \cite{{mysw99},{mamdo01}} and mass predictions
  \cite{{mysw96},{pear95}}, nuclei with  masses as large as A$\sim$300 could
  be formed.
  In this case the predicted abundance curve for nuclei below $A \sim 130 $ should
  also consider  triple (ternary) fission.     


 The probability of triple (or ternary)
fission \cite{{ternPere},{tripl}} is highly uncertain but for
nuclei with $A > 260$ near the neutron drip-line, fission into 3
fragments is energetically  possible and could occur. Present
\cite{tripl} and forthcoming \cite{new_exp} experiments should
clarify the probabilities of such a processes.

For the development of the r-process models a detailed analysis of
fission properties is required. Except for neutron-induced fission
and $\beta$-delayed fission, neutrino-induced fission was also
proposed recently \cite{Qian02}. In this connection one should
notice that neutrino rates \cite{LanMar02} can also have great
importance for specific astrophysical conditions. Precise fission
rates and mass fragment distributions after fission have to be
obtained and included in r-process calculations.


 After extended calculations of both beta-delayed fission
probabilities and neutron-induced fission rates for all nuclei
involved in the r-process, we hope to clarify the question of the
role of fission rates at different stages of nucleosynthesis for
more specific astrophysical conditions.
  Due to our new results (see \fref{bdfrates}), it seems that fission will affect
 the r-process, but further detailed studies are ahead.

\section*{Acknowledgements.} The authors thanks F.~G\"onnenwein,
S.~Goriely, C.~Wagemans and all colleagues, especially
participants of the Conferences: Nuclei in the Cosmos (Tokyo,
2002), Seminar on Fission (Pont d'Oye, 2003), Nucleus-2003 (Moscow
2003) for discussions on fission data, especially mass
distribution of fission fragments.

This work supported by grant 20-68031.02 of the Swiss National
Science Foundation and partially supported by grant 04-02-16793-a
of Russian Foundation of Basic Research (RFBR).

\end{document}